\documentclass[
 aip,
 jcp,
 amsmath,amssymb,
 reprint,
]{revtex4-1}

\usepackage{graphicx}
\usepackage{bm}                  
\usepackage[version=3]{mhchem}
\usepackage{textcomp}            
\usepackage{gensymb}             
\usepackage{float}
\usepackage[colorlinks,allcolors=black,citecolor=blue,urlcolor=blue]{hyperref}

\graphicspath{{figures/}}

\newcommand{\dd}{\mathrm{d}}
\newcommand{\vect}[1]{\mathbf{#1}}

\begin{document}

\title{{\it Ab initio} molecular dynamics with nuclear quantum effects at classical cost: ring polymer contraction for density functional theory}

\author{Ondrej Marsalek}
\author{Thomas E. Markland} \email{tmarkland@stanford.edu}
\affiliation{Department of Chemistry, Stanford University, Stanford, California 94305, USA}

\date{\today}

\begin{abstract}
Path integral molecular dynamics simulations, combined with an {\it ab initio} evaluation of interactions using electronic structure theory, incorporate the quantum mechanical nature of both the electrons and nuclei, which are essential to accurately describe systems containing light nuclei.
However, path integral simulations have traditionally required a computational cost around two orders of magnitude greater than treating the nuclei classically, making them prohibitively costly for most applications.
Here we show that the cost of path integral simulations can be dramatically reduced by extending our ring polymer contraction approach to {\it ab initio} molecular dynamics simulations.
By using density functional tight binding as a reference system, we show that our ring polymer contraction scheme gives rapid and systematic convergence to the full path integral density functional theory result.
We demonstrate the efficiency of this approach in {\it ab initio} simulations of liquid water and the reactive protonated and deprotonated water dimer systems.
We find that the vast majority of the nuclear quantum effects are accurately captured using contraction to just the ring polymer centroid, which requires the same number of density functional theory calculations as a classical simulation.
Combined with a multiple time step scheme using the same reference system, which allows the time step to be increased, this approach is as fast as a typical classical {\it ab initio} molecular dynamics simulation and 35$\times$ faster than a full path integral calculation, while still exactly including the quantum sampling of nuclei.
This development thus offers a route to routinely include nuclear quantum effects in {\it ab initio} molecular dynamics simulations at negligible computational cost.
\end{abstract}

\keywords{Path integral quantum mechanics, path integral molecular dynamics, multiple time scale molecular dynamics, ring polymer contraction, PIMD, RPMD, CMD, AIMD}

\maketitle

\section{Introduction}

Obtaining an accurate theoretical description of the atomistic properties of chemical and biological systems requires the development of {\it ab initio} simulation approaches that explicitly include the quantum mechanical nature of both the electrons and nuclei.
{\it Ab initio} molecular dynamics (AIMD) simulations, where the interactions are obtained from on-the-fly evaluation of the electronic structure, include the quantum nature of the electrons but treat the nuclei as classical particles.
Path integral molecular dynamics (PIMD) simulations allow the exact inclusion of nuclear quantum effects (NQEs) in static equilibrium (imaginary time) properties by exploiting the mapping of a quantum system onto an extended classical system of ring polymers comprised of multiple replicas (beads) of the classical system, with harmonic springs that link adjacent replicas of each atom in the system~\cite{Parrinello1984/10.1063/1.446740,Chandler1981/10.1063/1.441588,Feynman1965}.
By combining PIMD with AIMD, the resulting {\it ab initio} path integral molecular dynamics (AI-PIMD) simulations~\cite{Marx1994/10.1007/BF01312185,Marx1996/10.1063/1.471221,Tuckerman1996/10.1063/1.471771} thus allow one to capture the full interplay of nuclear and electronic quantum effects.

The imaginary time path integral formalism, which forms the basis of PIMD, also provides the foundation for a number of approximations to the quantum dynamics of chemical systems.
These range from methods that utilize imaginary time information to obtain their initial conditions~\cite{Liu2009/10.1063/1.3254372}, to those which analytically continue the imaginary time data~\cite{Habershon2007/10.1063/1.2786451,Rabani2002/10.1073/pnas.261540698}, to those that utilize the dynamics of the imaginary path itself to approximate quantum dynamics, such as ring polymer molecular dynamics (RPMD)~\cite{Craig2004/10.1063/1.1777575,Habershon2013/10.1146/annurev-physchem-040412-110122} and centroid molecular dynamics (CMD)~\cite{Cao1994/10.1063/1.468400}.

Despite their utility, PIMD and RPMD have traditionally required a considerable increase in computational cost compared to classical simulations.
In particular, the stiff equations of motion result in non-ergodic dynamics, which limits the time step that can be used and also the sampling efficiency\cite{Hall1984/10.1063/1.448112}.
This has led to the development of coordinate transformations and advanced thermostatting methods to alleviate these issues~\cite{Pollock1984/10.1103/PhysRevB.30.2555,Tuckerman1993/10.1063/1.465188,Ceriotti2010/10.1063/1.3489925,Ceriotti2011/10.1063/1.3556661,Ceriotti2012/10.1103/PhysRevLett.109.100604}.
However, even when these issues are addressed, another significant challenge remains: the need to create multiple replicas of the system.
Each of these replicas requires the evaluation of the potential energy and forces on all of its atoms.
Since obtaining the forces is the rate limiting step in AIMD calculations, the need to create the $P$ replicas makes the cost of an AI-PIMD simulation at least $P$ times higher than the corresponding AIMD simulation.
The number of replicas required depends on the highest frequency in the system, and for typical hydrogen-containing systems at room temperature is around 32 for most properties, but can rise as high as 128 if high accuracy of fluctuation properties, such as the heat capacity, is required~\cite{Shiga2005/10.1063/1.2035078,Shinoda2005/10.1103/PhysRevE.71.041204}.

A number of methods have been proposed to reduce the number of replicas required to converge AI-PIMD simulations.
Two particularly popular approaches are higher order factorization schemes~\cite{Perez2011/10.1063/1.3609120} and approaches based on the generalized Langevin equation (GLE)~\cite{Ceriotti2011/10.1063/1.3556661,Ceriotti2012/10.1103/PhysRevLett.109.100604}.
These methods are able to considerably reduce the number of replicas required, although both suffer from certain limitations.
The former requires the use of the Hessian in the calculation of atomic forces, making it impractical for AIMD, where obtaining the Hessian is very expensive.
This has led to the introduction of simulation schemes that evolve the system under the standard PIMD Hamiltonian and then utilize thermodynamic reweighting and higher order estimators to recover the rapid convergence with the number of replicas.
However, the efficiency of reweighting decreases with system size, limiting its use to small systems\cite{Ceriotti2011/10.1098/rspa.2011.0413}.
The latter approach of employing colored noise via GLE dynamics has been shown to greatly accelerate the convergence of properties, such as the quantum kinetic energy, in simulations ranging from aqueous systems~\cite{Wang2014/10.1063/1.4894287,Ceriotti2013/10.1073/pnas.1308560110} to enzymes~\cite{Wang2014/10.1073/pnas.1417923111}.
However, although noise matrices have been created that can be applied to a wide range of systems~\cite{GLE-input-generator}, the generation of new ones requires additional parameterization.
A given noise matrix will also only accelerate the convergence of the specific correlations of the imaginary time path, such as the virial kinetic energy estimator, that were targeted in the parameterization~\cite{Ceriotti2012/10.1103/PhysRevLett.109.100604,Ceriotti/10.1063/1.4772676}.
In addition, due to the non-equilibrium nature of the GLE dynamics, the probability distribution produced cannot be related to a well-defined ensemble, making its combination with Monte Carlo or reweighting unfeasible.
Finally, neither of these approaches yield dynamics that can be interpreted in the context of CMD or RPMD, limiting their use to static equilibrium properties.

A particularly appealing scheme that does not suffer from these deficiencies is ring polymer contraction (RPC)~\cite{Markland2008/10.1063/1.2953308}.
RPC employs a different approach: instead of reducing the total number of replicas, $P$, it reduces the cost of evaluating the forces on each of them.
To achieve this, RPC exploits the fact that the replicas are kept close in space due to the strong harmonic spring terms between them.
As a result, any smoothly varying interaction can be approximated with negligible error on a much coarser representation of the imaginary time path, i.e. one with fewer replicas.
For systems where the forces can be split into components that vary smoothly in space and those which vary rapidly, one can exploit this observation by evaluating the rapidly varying components on all replicas and the smoothly varying ones on a contracted ring polymer comprised of fewer replicas, $P'$.
If this splitting is constructed such that the computational cost of the rapidly varying forces is negligible compared to that of the smoothly varying forces, one can decrease the cost of the force evaluations by a factor of $P/P'$.
Despite the evaluation of some of the forces on the contracted ring polymer, these forces are exactly projected back onto the full ring polymer.
Thus at any instant the positions and forces on the full imaginary time path (ring polymer) representing each particle are known.

The RPC approach has previously been shown to be highly effective in reducing the cost of PIMD and RPMD simulations with empirical potentials by using separation of the inter- and intramolecular forces~\cite{Markland2008/10.1063/1.2953308}, by range separation of the Coulomb potential~\cite{Markland2008/10.1016/j.cplett.2008.09.019} and in polarizable force fields by splitting of the contributions to the polarization~\cite{Fanourgakis2009/10.1063/1.3216520}.
In these applications, the properties of interest converged systematically with the number of contracted replicas used and convergence was typically achieved with $P'$=1 to $P'$=6, allowing significant increases in efficiency and achieving near classical computational cost with $P'$=1.
In addition, RPC has a well defined Hamiltonian which can be used to assess integration accuracy and allows combination of RPC with reweighting or replica exchange techniques.
Finally, since RPC is simply an approximation to some of the interactions, it can be used to perform CMD and RPMD simulations from which approximate quantum dynamics can be obtained.
However, despite the success of these splittings for empirical force fields, the interactions obtained from electronic structure methods can not be trivially split into different types.

Here we show it is possible to extend RPC to AI-PIMD while retaining all its desirable features.
In particular, we demonstrate that our {\it ab initio} RPC (AI-RPC) scheme gives rapid convergence to the full AI-PIMD result obtained from density functional theory (DFT) for systems ranging from the reactive protonated and deprotonated water dimers to liquid water, by using self consistent charge density functional tight binding (SCC-DFTB)~\cite{Frauenheim2000} as a reference system.
We show that this convergence can usually be obtained with $P'=1$ contracted replicas for many properties, with full convergence by $P'=6$.
Further, by exploiting the smoothness in real as well as imaginary time generated by this choice of reference potential, we can increase the time step to 2~fs by using a multiple time scale (MTS) propagation scheme.
This combination of MTS and AI-RPC enables a further decrease of the computational cost of the method, yielding a scheme which is in many cases cheaper than a standard classical AIMD simulation while achieving an exact treatment of the NQEs.
This development thus opens the door to performing {\it ab initio} simulations that include NQEs routinely at negligible extra computational cost.

\section{Theory\label{sec:theory}}

In this section we briefly review the RPC formalism and outline the details necessary for its combination with {\it ab initio} evaluation of the interactions by using a reference force.

\subsection{Ring Polymer Contraction}

The Hamiltonian of a system of $N$ classical particles of masses $m_i$ is given by
\begin{equation}
H = \sum_{i=1}^N \frac{\vect{p}_i^2}{2m_i} + V(\vect{r}_1,\ldots,\vect{r}_N),
\end{equation}
where in AIMD the potential energy $V({\bf r}_1,\ldots,{\bf r}_N)$ is obtained from an electronic structure calculation.
The dynamics generated by this Hamiltonian gives rise to classical nuclear sampling of the electronic surface. For distinguishable particles the path integral expression for the quantum partition function corresponding to this Hamiltonian is
\begin{equation}
Z_P = \frac{1}{(2\pi\hbar)^{f}} \left(\prod_{i=1}^{N} \sqrt{\frac{m_{i}}{\tilde{m}_{i}}} \right) \int \dd^f{\bf p} \int \dd^f \vect{r} \
e^{-\frac{\beta}{P} H_P({\bf p}, \vect{r})},
\label{eq:Z_P}
\end{equation}
where $f=3NP$, $P$ is the number of replicas and $\beta=1/(k_{\rm B}T)$. $H_P(\vect{p},\vect{r})$ is the PIMD Hamiltonian,
\begin{equation}
\label{eq:H_P}
H_P({\bf p},{\bf r}) = H_0({\bf p},{\bf r}) + \sum_{j=1}^P V({\bf r}_1^{(j)},\ldots,{\bf r}_N^{(j)}),
\end{equation}
where
\begin{equation}
\label{eq:H_0}
H_0({\bf p},{\bf r}) =
\sum_{i=1}^N\sum_{j=1}^P \left( {\frac{|\vect{p}_{i}^{(j)}|^2}{2\tilde{m}_i}} +
                                \frac{1}{2}m_i\omega_P^2 |{\bf r}_{i}^{(j)}-{\bf r}_{i}^{(j-1)}|^2 \right)
\end{equation}
is the free ring polymer Hamiltonian, with cyclic boundary conditions, $ j + P \equiv j $, implied.
The dynamical (sampling) masses $\tilde{m}_i$ do not have to be the same as the physical masses $m_i$ and in PIMD simulations are typically set based on computational convenience.
We use the notation
\begin{equation}
\vect{p} \equiv \{\vect{p}_i^{(j)}\}_{i=1 \ldots N}^{j=1 \ldots P}, \: \vect{r} \equiv \{\vect{r}_i^{(j)}\}_{i=1 \ldots N}^{j=1 \ldots P}
\end{equation}
for the full set of ring polymer momenta and positions, respectively.
The Hamiltonian in Eq.~\ref{eq:H_P} corresponds to a set of $P$ copies of the physical system, where the adjacent replicas representing each particle are connected by harmonic springs of frequency $\omega_P=P/(\beta \hbar)$.
As $P \to \infty$, the classical dynamics generated by this Hamiltonian samples the exact quantum mechanical partition function. However, evaluating this Hamiltonian requires $P$ electronic structure calculations, to compute $V(\vect{r}^{(j)})$ for each of the $P$ replicas.

Let us now consider splitting the full force on a particle $\vect{f}_{\rm{full}}$, which requires a full electronic structure calculation to obtain, into a reference component $\vect{f}_{\rm{ref}}$ and the remaining difference force,
\begin{equation}
\label{eq:f-split}
\vect{f}_{\rm{diff}} = \vect{f}_{\rm{full}} - \vect{f}_{\rm{ref}}.
\end{equation}
If one constructs the reference system such that it captures the rapidly varying parts of the full interactions, the remaining difference force will be smoothly varying in space.
If this is the case, the smoothly varying force can be accurately approximated by evaluating it on a contracted ring polymer.
This is achieved by mapping the $P$-replica ring polymer positions, $\vect{r}$, onto a contracted set of positions, $\vect{r}' \equiv \{\vect{r}_i^{(j')}\}_{i=1 \ldots N}^{j'=1 \ldots P'}$, of $P'\le P$ replicas using the transformation
\begin{equation}
\label{eq:RPC-r}
\vect{r}_{i}^{(j')} = \sum_{j=1}^P T_{j'j}\,\vect{r}_{i}^{(j)}.
\end{equation}
A number of ways to achieve this transformation are possible.
Here we use our previously introduced normal mode contraction procedure, where the ring polymer is transformed to its local normal mode representation.
The $P-P'$ normal modes with the highest frequencies, which interact most weakly with the physical potential, are then discarded and the remaining $P'$ normal modes are transformed back to the coordinate representation \cite{Markland2008/10.1063/1.2953308}.
The required transformation matrix, $T_{j'j}$, is given in Ref.~\citenum{Markland2008/10.1063/1.2953308} and can be simply evaluated and applied at negligible computational cost.
In the limiting cases, the net effect of this transformation is such that when $P'=P$, it leaves the ring polymer unchanged and when $P'=1$ it contracts the ring polymer to its centroid, $\overline{\bf r}_i$,
\begin{equation}
\vect{r}_i^{(j'= P'=1)} = \overline{\bf r}_i = \frac{1}{P} \sum_{j=1}^P \vect{r}_i^{(j)}.
\end{equation}
For intermediate values, the transformation creates a contracted set of $P'$ positions which approximately represent the full ring polymer (effectively taking a lower-order Fourier representation of the imaginary time path).
The slowly varying difference force can then be evaluated on the contracted positions, $\vect{f}_{i,\rm{diff}}^{(j')}$, and then projected back onto the full ring polymer representing each particle:
\begin{equation}
\vect{f}_{i,\rm{diff}}^{(j)} = \frac{P}{P'} \sum_{j'=1}^{P'} T_{j'j} \vect{f}_{i,\rm{diff}}^{(j')},
\end{equation}
in which the factor of $P/P'$ on the right-hand side arises naturally from the ratio of the number of replicas in the full and contracted system~\cite{Markland2008/10.1063/1.2953308}.
With this contracted approximation to the difference force, the full force on each ring polymer is
\begin{equation}
\label{eq:RPC-f}
\begin{split}
\vect{f}_i^{(j)}
& = \vect{f}_{i,\rm{ref}}^{(j)} + \frac{P}{P'} \sum_{j'=1}^{P'} T_{j'j} \vect{f}_{i,\rm{diff}}^{(j')} \\
& = \vect{f}_{i,\rm{ref}}^{(j)} + \frac{P}{P'} \sum_{j'=1}^{P'} T_{j'j} \left( \vect{f}_{i,\rm{full}}^{(j')} - \vect{f}_{i,\rm{ref}}^{(j')} \right),
\end{split}
\end{equation}
where the second equality follows from the definition of the difference force (Eq.~\ref{eq:f-split}).
Hence the computationally expensive full force only needs to be evaluated $P'$ times for each configuration, rather than $P$ times as in a standard PIMD simulation.

Although the forces in Eq.~\ref{eq:RPC-f} are an approximation to the full forces on each replica, the dynamics generated from them formally conserve a well defined Hamiltonian~\cite{Markland2008/10.1063/1.2953308}
\begin{equation}
\label{eq:RPC-v}
H'_P({\bf p},{\bf r}) = H_0({\bf p},{\bf r}) + \sum_{j=1}^P V_{\rm{ref}}^{(j)}(\vect{r}) + \frac{P}{P'} \sum_{j'=1}^{P'} V_{\rm{diff}}^{(j')}(\vect{r}'),
\end{equation}
where $V_{\rm{ref}}$ and $V_{\rm{diff}}$ are the potential energies that correspond to the reference and difference forces, respectively, and together replace the potential energy term in Eq.~\ref{eq:H_P}.
The existence of this Hamiltonian allows energy conservation to be checked during dynamics to assess integration accuracy and also enables the combination of RPC with reweighting schemes and methods that require a well defined ensemble.

It is important to note that since the contracted forces are projected back onto the full ring polymers which are then evolved in time under these forces, the positions of all the replicas that comprise full ring polymer are known.
These $P$ replica positions can therefore be used to evaluate any position dependent property at each time step.
For both forces and potential energies, one naturally generates the contracted estimators for them during the RPC evolution (Eqs.~\ref{eq:RPC-f} and \ref{eq:RPC-v}).
The forces on each replica obtained in a PIMD simulation can be used to compute the quantum kinetic energy of a particle, $i$, as the ensemble average of the virial estimator~\cite{Herman1982/10.1063/1.442815},
\begin{equation}
\mathcal{T}_{i} (\vect{r}) = \frac{3 k_{\rm{B}} T}{2} - \frac{1}{2P}\sum_{j=1}^P (\vect{r}_i^{(j)}-\overline{\bf r}_i)\cdot \vect{f}^{(j)}_{i}.
\end{equation}
However, for these properties an alternative approach to achieve more accurate results from a RPC trajectory is to use uncontracted estimators (UE)\cite{Markland2008/10.1063/1.2953308}.
These UEs are evaluated by performing full (uncontracted) potential and force evaluations on the full $P$ replica configurations generated from the contracted simulation.
This post-processing of the RPC configurations requires extra full potential and force evaluations and hence some additional cost.
However, by choosing to only post-process a subset of frames, one can take advantage of the fact that adjacent time steps are heavily correlated, and hence evaluating the UE adds comparatively little computational overhead.

\subsection{Constructing a reference system\label{subsec:reference}}

RPC provides a way to save computational effort by using a physically motivated approximation in the evaluation of physical interactions in the system while still retaining the full dimensionality of the $P$-replica imaginary time path.
The form of our contraction transformation, Eq.~\ref{eq:RPC-r}, and the definition of the difference force, Eq.~\ref{eq:f-split}, guarantee convergence as $P'\to P$.
However, the rate of this convergence and the computational cost savings obtained from reducing the number of replicas in the contracted ring polymer depend crucially on the creation of an efficient reference system that accurately captures the rapidly varying parts of the interactions in the full system and thus leaves a slowly varying difference force.

The objective is therefore to construct a reference force $f_{\rm{ref}}$ which, when taken from the full electronic structure force $f_{\rm{full}}$, leaves a remaining force $f_{\rm{diff}}$ that can be well approximated on the contracted polymer.
The reference force for an AI-RPC approach should thus satisfy the following requirements:
\begin{enumerate}
\item It is computationally quick to evaluate compared to performing the full electronic structure calculation (ideally by at least a factor of 10). 
\item It gives a difference force that is smoothly varying in space.
\item It does not make assumptions on chemical bonding, i.e. it allows bond breaking and formation as the simulation progresses.
\end{enumerate}

It is vital to note that the reference force only has to leave a slowly varying remainder --- i.e. the reference force can be something that would give a very poor result for the dynamics and structure of the system if used alone (without the difference force which corrects for its deficiencies).
For example, when applied to empirical potentials, RPC has previously been shown to be successful even when the reference force was chosen as just the intramolecular force~\cite{Markland2008/10.1063/1.2953308}.
Hence, used alone, this reference system would predict no interactions between molecules (an ideal gas).
However, the presence of the difference force on the contracted replicas corrects for this, and it was shown~\cite{Markland2008/10.1063/1.2953308} that convergence of many structural properties of liquid water was obtained even with $P'=1$.

The idea of a reference system has natural origins in the MTS molecular dynamics community, where the reversible reference systems propagator algorithm (r-RESPA)~\cite{Tuckerman1992/10.1063/1.463137} was formulated as a method to allow efficient time-reversible symplectic propagation in molecular dynamics simulations with multiple components of the force which vary on different time scales.
Whereas MTS schemes exploit the slowly varying nature of some forces in real time to extend the propagation time step, RPC takes advantage of the spatially smooth variation of the forces in the imaginary time of the path integral.
Thus, the considerations needed for a good reference force in the two approaches are similar. This suggests that for systems where a suitable reference force can be identified, it can be used to naturally utilize both MTS and RPC in a simulation.

\begin{figure}[tbh]
\includegraphics[width=\columnwidth]{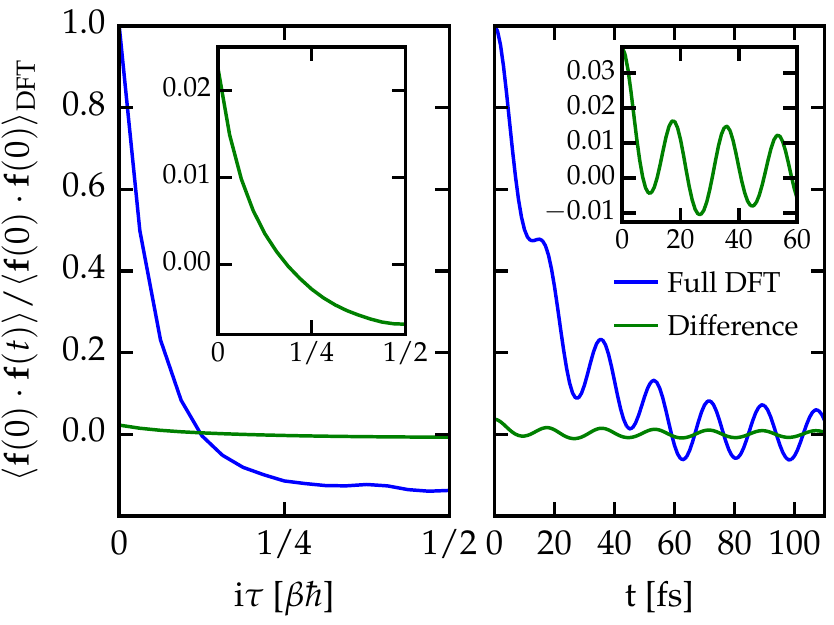}
\caption{
\label{fig:FFCF}
Force autocorrelation functions along the imaginary time path (left panel) and in real time (right panel) for the protons in the protonated water dimer from a 32 replica PIMD simulation.
The correlation function of the full DFT forces is shown in blue, while the correlation function of the difference between the DFT and DFTB forces is shown in green.
Correlation functions are scaled relative to the full DFT result at $t$=0 and i$\tau$=0, $\langle \vect{f}(0) \cdot \vect{f}(0) \rangle_{\rm{DFT}}$.
The insets show the details of the difference force.
}
\end{figure}

This complementarity is illustrated in Fig.~\ref{fig:FFCF}, which shows the force autocorrelation function of all the protons in the protonated water dimer along the imaginary time path (left panel) and in real time (right panel) obtained from a 32 replica PIMD simulation.
The full DFT force (blue line) decays rapidly along the imaginary time path, which extends from $i\tau = 0$ to $\beta \hbar$ (in the language of the discretized ring polymer, from replica 1 to 32 in this case).
Due to the cyclic boundary conditions on the path, the furthest distance in imaginary time between the replicas is at i$\tau = \beta \hbar / 2$ (i.e. the force correlation between replica 1 and replica 16). 
Likewise, in real time the full force also rapidly decays and additionally exhibits oscillations arising from bond vibrations. 

We now consider using SCC-DFTB as the reference force. The green line in Fig.~\ref{fig:FFCF} shows the difference force which would be obtained using this choice. In both real and imaginary time, the difference force is roughly 2 orders of magnitude smaller and is also much more smoothly varying than the full force. This suggests that the reference force captures almost all of the variation of the full force both along the ring polymer as well as in real time.
Hence the SCC-DFTB reference force is able to provide a smooth and small difference force in both real and imaginary time. In addition, using a SCC-DFTB reference for DFT also satisfies the other desirable features such as the ability to make and break bonds and computational savings of typically 2 orders of magnitude over DFT due to the tabulation of the Hamiltonian and overlap matrix elements and the use of a minimal valence basis set\cite{Gaus2011/10.1021/ct100684s}.

One could also imagine using a number of other methods that have been suggested to split {\it ab initio} interactions for MTS schemes to extend RPC to {\it ab initio} PIMD. For DFT, these range from using a reference force generated by using reduced basis sets and non-iterative Harris functionals \cite{Anglada2003/10.1103/PhysRevE.68.055701}, neglecting Hartree-Fock exchange in hybrid functionals \cite{Guidon2008/10.1063/1.2931945}, range seperation of the Coulomb operator \cite{Luehr2014/10.1063/1.4866176} or by using a reference empirical force field \cite{Geng2015/10.1016/j.jcp.2014.12.007,Luehr2014/10.1063/1.4866176}. For higher level wavefunction methods, such as MP2, reference forces that have been suggested include neglecting the dynamic electron correlation \cite{Steele2013/10.1063/1.4812568}, aggressive electron integral screening \cite{Fatehi2015/10.1021/ct500904x}, and using DFT as a reference for the higher level theory~\cite{DelBen2015/10.1063/1.4927325}.

\subsection{Multiple time scale integration}

Given the natural complementarity of MTS, which exploits the smoothness of the difference force in real time, and RPC, which exploits its smoothness in imaginary time, we combine both approaches in our simulations.
Since the utility and theory behind MTS approaches are well established and that we have previously combined RPC and MTS a number of times for empirical potentials\cite{Habershon2009/10.1063/1.3167790,Markland2011/10.1038/nphys1865,Markland2012/10.1063/1.3684881,Markland2012/10.1073/pnas.1203365109}, here we only sketch the relevant details specific to our combination with AI-RPC.

The MTS integration used here was given by the factorization of the propagator
\begin{equation}
\begin{split}
\exp\left(i\mathcal{L} \Delta t \right) = \\
& \hspace{-1.75cm} \exp\left(i\mathcal{L}_\gamma \Delta t / 2 \right)
\exp\left(i\mathcal{L}_{\rm{diff}} \Delta t / 2 \right) \times \\
& \hspace{-1.75cm} \left[
\exp\left(i\mathcal{L}_{\rm{ref}} \delta t / 2 \right)
\exp\left(i\mathcal{L}_{\rm{0}} \delta t \right)
\exp\left(i\mathcal{L}_{\rm{ref}} \delta t / 2 \right)
\right]^M \times \\
& \hspace{-1.75cm} \exp\left(i\mathcal{L}_{\rm{diff}} \Delta t / 2 \right)
\exp\left(i\mathcal{L}_\gamma \Delta t / 2 \right),
\end{split}
\end{equation}
where $\Delta t = M \delta t$ is the outer time step and $\delta t$ is the inner time step. Here the propagator $\exp\left(i\mathcal{L}_\gamma \Delta t / 2 \right)$ evolves the system under the thermostat for the time interval $\Delta t/2$ and $\exp\left(i\mathcal{L}_{\rm{diff}} \Delta t / 2 \right)$ evolves the system momenta under the difference force by $\Delta t/2$. In the inner loop, $\exp\left(i\mathcal{L}_{\rm{ref}} \delta t / 2 \right)$  evolves the system momenta under the reference force by half a small time step $\delta t / 2$ and $\exp\left(i\mathcal{L}_{\rm{0}} \delta t \right)$ evolves the system under the influence of the free ring polymer Hamiltonian $H_0$ by transforming to the free ring polymer normal mode representation, which allows analytic integration.
The explicit operations of each of these propagators are given in Ref.~\citenum{Ceriotti2010/10.1063/1.3489925}.

One major consideration in MTS schemes, which does not arise in RPC, is the existence of the resonance barrier, which limits the largest outer time step which can be used~\cite{Han2007/10.1016/j.cpc.2006.10.005}.
The first resonance can be shown to occur when the outer time step exceeds $\Delta t_{max} = \tau/\pi$, where $\tau$ is the time-period of the fastest mode in the problem~\cite{Han2007/10.1016/j.cpc.2006.10.005}.
For liquid water, where the highest physical frequency is the O-H stretch at $\sim$3600~cm$^{-1}$, this yields a value of $\Delta t_{max}$=2.95~fs.
However, the harmonic springs linking the replicas in PIMD give rise to additional
high-frequency modes.
The highest ring polymer normal mode frequency for the free ring polymer is~\cite{Ceriotti2010/10.1063/1.3489925}
\begin{equation}
\omega_{\rm{RP,max}} = \frac{2 P}{\beta \hbar}
\end{equation}
and for a ring polymer in a physical harmonic potential of frequency $\omega$ it shifts to
\begin{equation}
\sqrt{\omega_{\rm{RP,max}}^{2}+\omega^2}.
\end{equation}
For liquid water with $P=32$ at $T$=300~K, the highest frequency in the PIMD simulation is therefore $\sim$13345~cm$^{-1}$ and hence the first resonance would be expected at $\Delta t$=0.8~fs. To allow us to avoid the resonance barrier and thus use larger outer time steps in our MTS simulations, we use the standard procedure of
shifting all the normal mode frequencies to a single sufficiently low frequency by adjusting the normal modes' dynamical masses~\cite{Tuckerman1993/10.1063/1.465188,Cao1994/10.1063/1.468400,Hone2006/10.1063/1.2186636}.
In cases where changing the masses is not an option, one could consider coupling the highest normal modes to a strong thermostat to surpass the resonance barrier using either colored noise~\cite{Morrone2011/10.1063/1.3518369} or targeted white noise~\cite{Ceriotti2010/10.1063/1.3489925}.
The PILE thermostat scheme achieves the latter by transforming to the normal mode representation and coupling each normal mode to a Langevin thermostat with critical damping based on its free ring polymer frequency.
One case where the masses of the normal modes are determined and should not be changed is when obtaining approximate quantum dynamics using the RPMD approach where, $\tilde{m}_i = m_i$.
In this case, the TRPMD method~\cite{Rossi2014/10.1063/1.4883861}, which uses the PILE thermostat with a specific choice of damping coefficients to obtain dynamics, might be advantageous, since it also has damping applied to the normal modes of the ring polymer.
Indeed, results along these lines do suggest that one may be able to go beyond the first resonance barrier when the PILE equations of motion are used~\cite{Habershon2009/10.1063/1.3167790,Markland2012/10.1073/pnas.1203365109,Kapil2015/10.1063/1.4941091}.

\section{Simulation details}

We performed AI-PIMD and AIMD and AI-RPC simulations of the gas-phase protonated and deprotonated water dimer, and of bulk liquid water at T=300~K.
The liquid water simulations were performed under NVT conditions using a system of 64 water molecules with periodic boundary conditions.
A cubic box with sides of length 12.42~\AA\ was used, giving a density of 1000.8 kg/m$^3$.

The benchmark quantum simulation of liquid water used for comparison was performed using the PIGLET scheme~\cite{Ceriotti2012/10.1103/PhysRevLett.109.100604}.
Due to the accelerated convergence with the number of replicas afforded by PIGLET, we use 6 replicas, which has previously been shown to give excellent agreement for the position based properties and quantum kinetic energies to the exact path integral result~\cite{Ceriotti2012/10.1103/PhysRevLett.109.100604}.

In all other benchmark PIMD and in the AI-RPC simulations, the total number of replicas was set to $P$=32 and the dynamical masses of all ring polymer normal modes were rescaled to shift their frequencies to 500~cm$^{-1}$ to ensure accurate integration of the equations of motion\cite{Tuckerman1993/10.1063/1.465188}.
The inner time step was $\delta t$=0.5~fs in all simulations and in trajectories run with MTS, the outer time step was $\Delta t$=2.0~fs ($M$=4).
The PILE-G thermostat~\cite{Ceriotti2010/10.1063/1.3489925} was used to sample the canonical ensemble.
In this thermostatting scheme, a critical damping Langevin thermostat is attached to each ring polymer normal mode of each particle, while the total kinetic energy of the particle centroids is coupled to a single stochastic velocity rescaling thermostat.
In our classical MD simulations, the temperature was maintained using a single global stochastic velocity rescaling thermostat~\cite{Bussi2007/10.1063/1.2408420}.
The i-PI program~\cite{Ceriotti2013/10.1016/j.cpc.2013.10.027}, including the implementation of MTS~\cite{Kapil2015/10.1063/1.4941091}, was used to perform all the molecular dynamics simulations.
Interactions were obtained by calling external electronic structure programs through the socket interface.

For the gas-phase protonated and deprotonated dimers, DFT interactions were evaluated using the B3LYP hybrid density functional~\cite{Becke1993/10.1063/1.464913} and the 6-311++G** basis set in the Gaussian program~\cite{g09} called through the calculator provided by the Atomic Simulation Environment~\cite{Bahn2002/10.1109/5992.998641}.
SCC-DFTB~\cite{Frauenheim2000} interactions including diagonal DFTB3 terms~\cite{Gaus2011/10.1021/ct100684s} were evaluated in CP2K~\cite{Hutter2014/10.1002/wcms.1159} using the provided parameterization for O and H atoms.

In our liquid water simulations both the DFT and SCC-DFTB interactions were evaluated using the CP2K program~\cite{Vandevondele2005/10.1016/j.cpc.2004.12.014,Hutter2014/10.1002/wcms.1159}.
For the DFT interactions, we used the revPBE generalized gradient approximation (GGA) density functional~\cite{Perdew1996/10.1103/PhysRevLett.77.3865,Zhang1998/10.1103/PhysRevLett.80.890} with the PBE DFT-D3 dispersion correction\cite{Grimme2010/10.1063/1.3382344}.
Atomic cores were represented using the dual-space Goedecker-Tetter-Hutter pseudopotentials~\cite{Goedecker1996/10.1103/PhysRevB.54.1703}.
Kohn-Sham orbitals were expanded in a double $\zeta$ contracted Gaussian atomic basis set with polarization functions (DZVP) and a cutoff of 400~Ry was used for the auxiliary plane-wave basis set of the GPW method~\cite{Lippert1997/10.1080/002689797170220}.
The self-consistent field cycle was converged to an electronic gradient tolerance of $\epsilon_{\rm{SCF}} = 5 \times 10^{-7}$ using the orbital transformation method~\cite{VandeVondele2003/10.1063/1.1543154} with the initial guess provided by the always stable predictor-corrector extrapolation method~\cite{Kolafa2004/10.1002/jcc.10385,Richters2014/10.1063/1.4869865} at each molecular dynamics step.
SCC-DFTB interactions~\cite{Frauenheim2000}, including diagonal DFTB3 terms~\cite{Gaus2011/10.1021/ct100684s}, were combined with the D3 dispersion correction~\cite{Grimme2010/10.1063/1.3382344} and evaluated in periodic boundary conditions, using Ewald summation for electrostatics.

For the protonated gas phase dimer, the benchmark quantum trajectory was run for 125~ps and each of the contracted simulations for at least 250~ps.
For the deprotonated dimer, the benchmark trajectory was run for 120~ps and each contracted simulation for at least 300~ps.
For liquid water, the benchmark quantum trajectory was 80~ps and all other simulations were run for 100~ps.
Uncontracted estimators were evaluated by post-processing 100~ps of the trajectory with configurations taken every 4~fs for the protonated and deprotonated dimers and every 10~fs for liquid water.
Statistical error estimates on all the quantities reported were calculated using the bootstrapping method as 99~\% confidence intervals~\cite{Young2012/10.1007/978-3-319-19051-8}.

\section{Results}

To assess the accuracy and efficiency of our AI-RPC scheme, we performed simulations of the protonated and deprotonated water dimers in the gas phase and of bulk liquid water.
The former two were chosen for the reactive nature of the proton defects, which allow us to test the ability of our approach to capture on-the-fly making and breaking of chemical bonds. Liquid water was chosen to demonstrate the applicability of our method to a condensed-phase system that exhibits a delicate balance between its NQEs.
In particular, upon including NQEs water exhibits increased hydrogen bond deformation, which acts to weaken the hydrogen bonded network and destructure the liquid, and also increased proton sharing in the hydrogen bond, which has the opposite effect \cite{Habershon2009/10.1063/1.3167790}.
Liquid water thus allows us to assess if these competing quantum effects are correctly captured by AI-RPC.

\subsection{Protonated water dimer}

We first consider the protonated water dimer (an excess proton shared between two water molecules).
\begin{figure}[!b]
\includegraphics[width=\columnwidth]{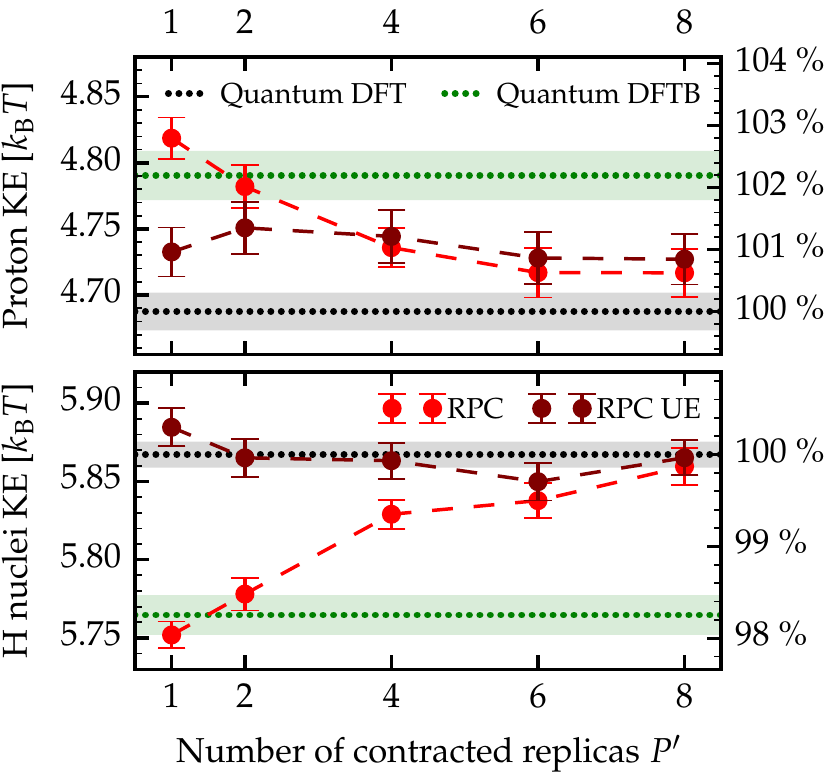}
\caption{
Convergence of the proton kinetic energy with respect to the number of contracted replicas $P'$ for the protonated water dimer.
Kinetic energies from trajectories run with a fixed total number of replicas $P$=32 and an increasing number of contracted replicas $P'$=1 to $P'$=8 are shown in red.
The values obtained using the uncontracted quantum kinetic energy estimator (UE) evaluated on geometries from the corresponding trajectories are shown in dark red.
The top panel shows data for the dangling protons, while the bottom panel shows data for the shared proton.
The quantum DFT and DFTB results, shown as dotted lines, were obtained from full 32-replica PIMD simulations.
The axis on the right shows the percentage of the total kinetic energy obtained relative to the quantum DFT result.
The classical contribution to the kinetic energy for each proton is 1.5 $k_{\rm{B}}T$.
Statistical error estimates are indicated using error bars and shading.
\label{fig:PD-ke}
}
\end{figure}
In order to test the systematic convergence of our contraction scheme, Fig.~\ref{fig:PD-ke} shows the quantum kinetic energy of the shared proton (top panel) and the dangling hydrogen nuclei (bottom panel) as the number of contracted replicas, $P'$, is increased.
Due to the protons' different chemical environments, the quantum kinetic energy of the weakly bound shared proton is 25~\% less than that of the hydrogen nuclei, which are covalently bound to the water molecules.
Already for a 1-replica contraction, which contracts the ring polymer to its centroid for the evaluation of the DFT interactions, both kinetic energies are within 3~\% of the full quantum DFT result.
This corresponds to an error of only 0.1~$k_{\rm{B}}T$ and is therefore already sufficiently converged for all practical purposes, and indeed is within the statistical sampling error of most AI-PIMD simulations.
As the number of contracted replicas is increased, the result is observed to systematically converge to the exact quantum result by $P'=8$, allowing high precision to be obtained by increasing the contraction level. In contrast, a standard PIMD simulation with 8 replicas gives an error in the kinetic energy of $\sim$23~\%. 
Using uncontracted estimators (UE) to post-process the trajectories brings the error of the 1-replica contraction to below 0.05~$k_{\rm{B}}T$.
In all cases, the oxygen nuclei kinetic energies were found to be within 0.01~$k_{\rm{B}}T$ of the full quantum DFT result, which is within the statistical error bars of our simulations. In addition, using MTS to increase the outer time step to 2~fs gave almost identical kinetic energies for $P'=1$ and $P'=6$ as well as for the UE for $P'=1$ (not shown in the plot since the points overlap with those shown), compared to the 0.5~fs simulations, suggesting that the combination with MTS introduces negligible error.

\begin{figure}[b]
\includegraphics[width=\columnwidth]{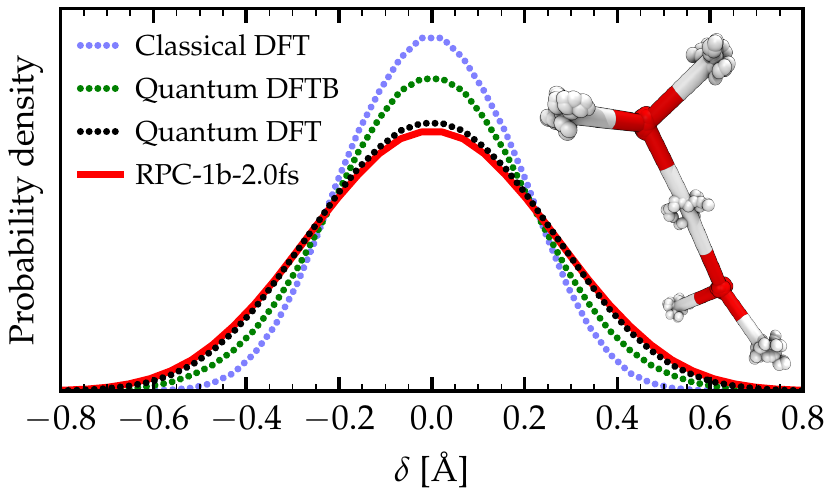}
\caption{
Probability densities along the proton transfer coordinate, $\delta$, for the protonated water dimer at 300~K.
The quantum DFT and DFTB results were obtained from 32-replica PIMD simulations.
\label{fig:PD-delta}
}
\end{figure}

\begin{figure}[h]
\includegraphics[width=\columnwidth]{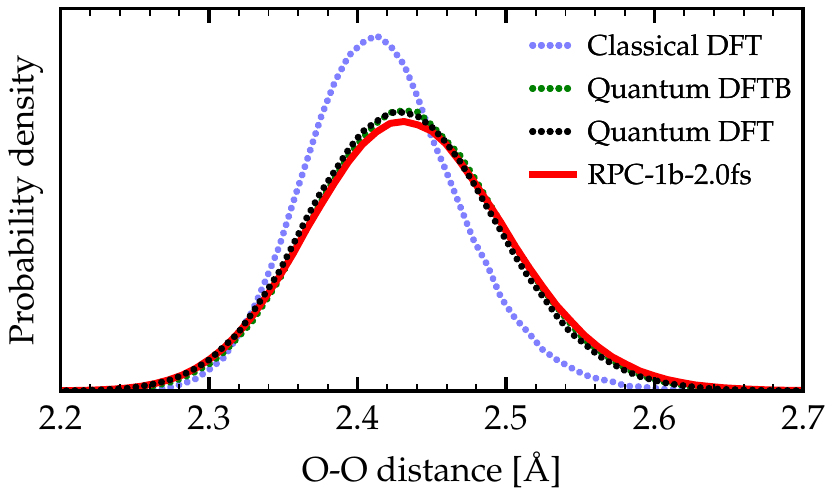}
\caption{
Probability densities of the O-O distance for the protonated water dimer at 300~K.
The quantum DFT and DFTB results were obtained from 32-replica PIMD simulations.
\label{fig:PD-OO}
}
\end{figure}

To characterize the position of a proton between two oxygen atoms, we use the proton sharing coordinate $\delta$ defined as
\begin{equation}
\delta = d_{\rm{OH}} - d_{\rm{O'H}},
\end{equation}
where $d_{\rm{OH}}$ and $d_{\rm{O'H}}$ are the distances of the proton from the oxygen atoms.
Fig.~\ref{fig:PD-delta} shows the probability distribution along the $\delta$ coordinate.
Here a contraction to the centroid combined with MTS using a 2~fs outer time step (RPC-1b-2.0fs) is sufficient to capture the broader distribution of proton positions within the statistical error bars, which is not captured by either classical DFT or the quantum simulation of the pure DFTB system.
This wider distribution of the shared proton positions upon including NQEs changes the strength of the binding to the water molecules and thus also changes the O-O distance distribution shown in Fig.~\ref{fig:PD-OO}.
Again, AI-RPC is in excellent agreement with the full quantum DFT results, even with a centroid contraction. The O-O distribution is also captured almost exactly by a pure quantum DFTB simulation.
However, since the quantum DFTB simulation does not obtain the correct quantum proton $\delta$ distribution, this agreement is likely fortuitous.

\subsection{Deprotonated water dimer}

The deprotonated water dimer provides a particularly interesting test case, as NQEs cause a qualitative change in the behavior of the shared proton~\cite{Tuckerman1997/10.1126/science.275.5301.817}.
With classical nuclei, a double peak in the distribution of the proton sharing coordinate $\delta$ is observed, i.e. it exists primarily as a hydroxide ion bound to one of the water molecules, \ce{H2O-OH-}, with frequent switching in which oxygen atom holds the proton defect.
However, upon including NQEs this distribution changes to a single broad peak i.e. corresponding to a situation that more closely resembles a delocalized proton bound between two hydroxide like species, \ce{[HO-H-OH]-}.
This presents a more challenging test case, since the DFTB reference system is a much poorer approximation to DFT for this problem.
For example, in contrast to DFT, a classical DFTB simulation already predicts that the proton should be symmetrically tightly shared even without NQEs included.
\begin{figure}[t]
\includegraphics[width=\columnwidth]{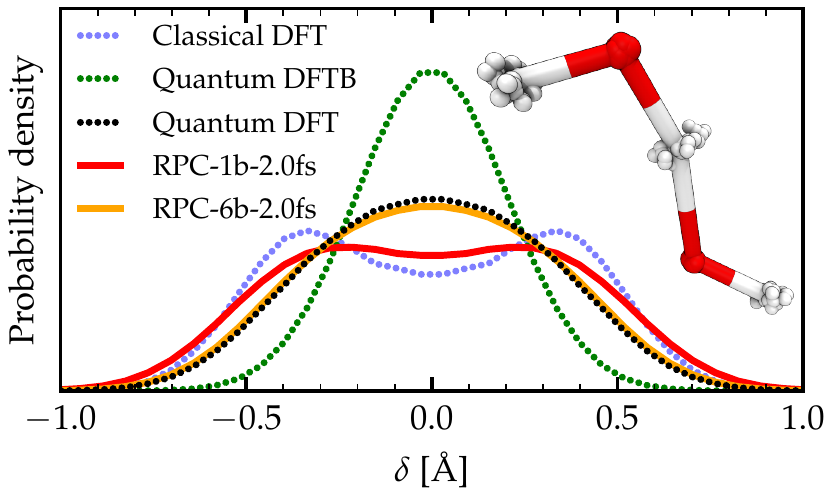}
\caption{
Probability densities along the proton transfer coordinate, $\delta$, for the deprotonated water dimer at 300~K.
The quantum DFT and DFTB results were obtained from 32-replica PIMD simulations.
\label{fig:DD-delta}
}
\end{figure}
\begin{figure}[H]
\includegraphics[width=\columnwidth]{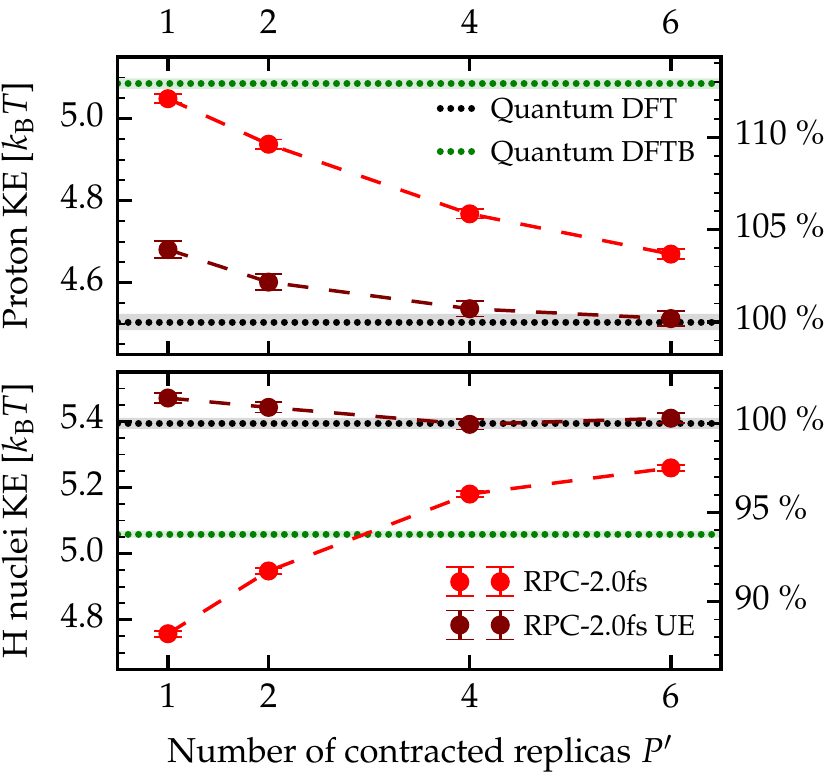}
\caption{
Convergence of the proton kinetic energy with respect to the number of contracted beads $P'$ for the deprotonated water dimer.
Kinetic energies from trajectories run with a fixed total number of replicas $P$=32 and the number of contracted replicas from $P'$=1 to $P'$=6 and multiple time stepping ($M$=4, $\Delta t$=2.0~fs) are shown in red.
The values obtained using the uncontracted quantum kinetic energy estimator (UE) evaluated on geometries from the corresponding trajectories are shown in dark red.
The top panel shows data for the shared proton and the bottom panels shows the data for the dangling H nuclei.
The quantum DFT and DFTB results, shown as dotted lines, were obtained from full 32-replica PIMD simulations.
The axis on the right shows the percentage of the total kinetic energy obtained relative to the quantum DFT result.
The classical contribution to the kinetic energy for each proton is 1.5~$k_{\rm{B}}T$.
The shading around the lines and error bars show the statistical error estimates.
\label{fig:DD-ke}
}
\end{figure}
The degradation in the reference system leads to a centroid contracted result (RPC-1b-2.0fs) for the proton $\delta$ distribution, shown in Fig.~\ref{fig:DD-delta}, which does not capture all of the quantum DFT result and still retains a remnant of the classical-like double peak structure.
However, even with the poorer reference, increasing the number of contracted replicas to $P'=6$ allows good agreement with the quantum DFT results to be obtained, again highlighting the systematic convergence of AI-RPC.

This slower convergence with the number of contracted replicas due to the poorer performance of the DFTB reference also manifests strongly in the quantum kinetic energies in Fig.~\ref{fig:DD-ke}, where a 1-replica contraction gives an error of 0.54~$k_{\rm{B}}T$ (11.9~\%) for the shared proton.
This error decreases to a much more acceptable value of 0.16~$k_{\rm{B}}T$ (3.5~\%) when 6 contracted replicas are used, which still gives a considerable reduction in computational cost compared to a full quantum DFT calculation.
In addition, using UEs reduces the error in the kinetic energy from a 1-replica contraction to within 0.17~$k_{\rm{B}}T$ (3.7~\%) and 0.08~$k_{\rm{B}}T$ (1.5~\%) of the exact result for the proton and H nuclei, respectively.
Using UEs with 4 contracted replicas, gives kinetic energies that are within the error bars of the full quantum DFT result.

\subsection{Liquid water}

\begin{figure}[h]
\includegraphics[width=\columnwidth]{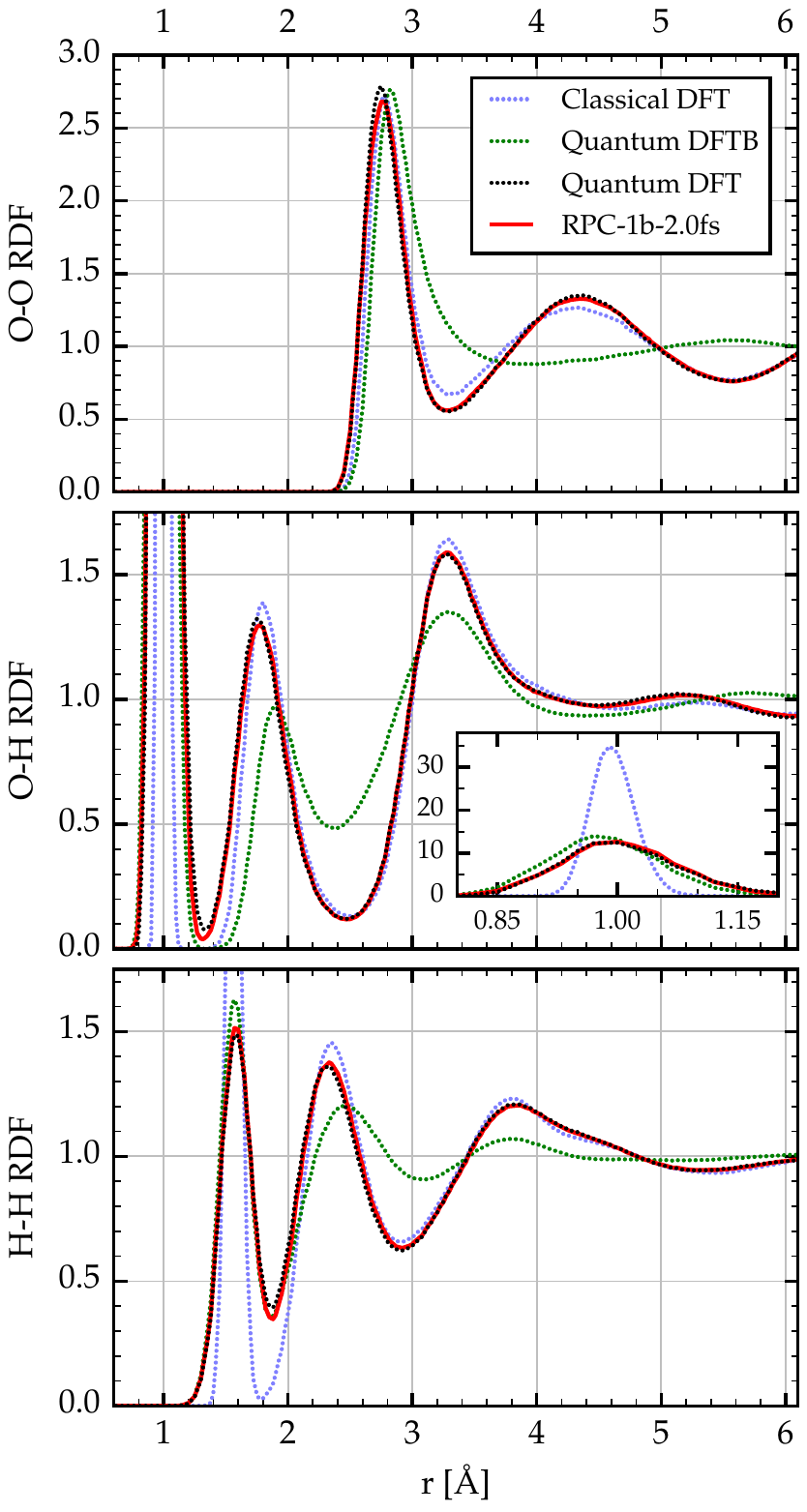}
\caption{
Radial distribution functions of liquid water at 300~K obtained from AI-RPC and comparison simulations.
The quantum DFT results were obtained from a 6-replica PIGLET simulation, while the quantum DFTB results were obtained using a 32-replica PIMD simulation.
A 1-replica contraction with a 2~fs outer time step (RPC-1b-2.0fs) is observed to give results graphically identical to the full quantum DFT result.
\label{fig:water-rdf}
}
\end{figure}

Owing to the tight binding approximation in DFTB, the method generally performs better for covalently bonded systems, and shows notable deficiencies in describing the structure of weakly bound and hydrogen bonded systems\cite{Koskinen2009/10.1016/j.commatsci.2009.07.013,Maupin2010/10.1021/jp1010555,Goyal2013/10.1021/jp503372v}.
This problem is apparent for liquid water, where, as shown in top panel of Fig.~\ref{fig:water-rdf}, the quantum DFTB O-O radial distribution function (RDF) lacks any structure beyond the first peak\cite{Hu2007/10.1021/jp070308d}.
From this, one might expect that an AI-RPC scheme based on DFTB would converge very slowly for liquid water as the number of contracted replicas is increased.
However, as shown in Fig.~\ref{fig:water-rdf}, AI-RPC with contraction to a single replica yields RDFs that are graphically identical to full quantum DFT, thus further demonstrating the robustness of our approach.

The reason for the rapid convergence can be elucidated by examining the O-H and H-H RDFs obtained from pure DFTB, where the first (intramolecular) peaks show good agreement, while the remaining ones are noticeably shifted and broadened compared to DFT.
As highlighted in Sec.~\ref{subsec:reference}, the main requirement for a good choice of the reference force for our AI-RPC scheme is that it yields a smoothly varying difference force when subtracted from the full DFT force.
Hence, while DFTB does not capture the smoothly varying interactions at longer ranges in DFT correctly, these are corrected by the full DFT interactions, evaluated on the contracted replicas.

\begin{figure}[h]
\includegraphics[width=\columnwidth]{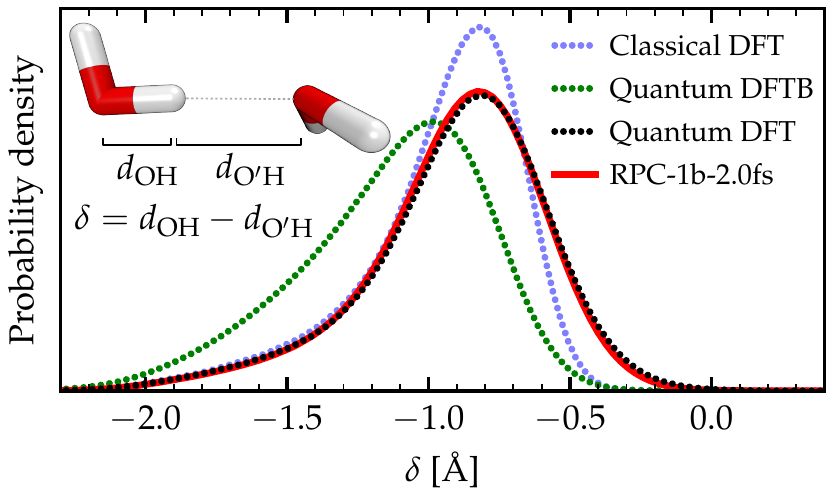}
\caption{
Distribution of the proton transfer coordinate, $\delta$, in liquid water at 300~K from AI-RPC and comparison simulations.
The quantum DFT results shown obtained from a 6-replica PIGLET simulation, while the quantum DFTB results were obtained using a 32-replica PIMD simulation.
A 1-replica contraction with a 2~fs outer time step (RPC-1b-2.0fs) is observed to give results graphically identical to the full quantum DFT result.
The atom distances are labeled in a snapshot of a hydrogen bond.
\label{fig:water-delta}
}
\end{figure}

\begin{figure}[h]
\includegraphics[width=\columnwidth]{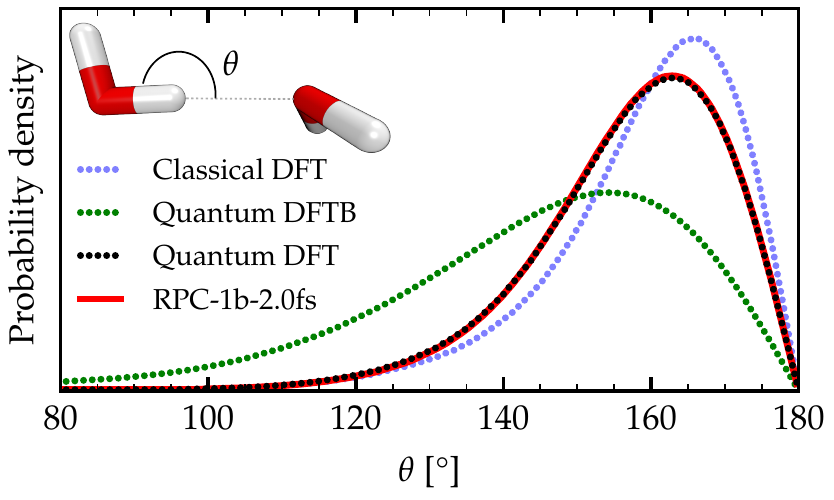}
\caption{
\label{fig:water-theta}
Distribution of the hydrogen bond angle, $\theta$, in liquid water at 300~K from AI-RPC and comparison simulations.
The quantum DFT results shown obtained from a 6-replica PIGLET simulation, while the quantum DFTB results were obtained using a 32-replica PIMD simulation.
A 1-replica contraction with a 2~fs outer time step (RPC-1b-2.0fs) is observed to give results graphically identical to the full quantum DFT result.
The angle $\theta$ is labeled in a snapshot of a hydrogen bond.
}
\end{figure}

The accuracy of AI-RPC using a DFTB reference force is also apparent in the rapid convergence of the local geometries of the hydrogen bonds in water engendered in the distribution of the proton sharing coordinate $\delta$ in Fig.~\ref{fig:water-delta} and in the distribution of hydrogen bond angles $\theta$ in Fig.~\ref{fig:water-theta}.
These distributions, which show large quantum effects, describe the delocalization of protons in hydrogen bonds in the direction along the hydrogen bond ($\delta$) and perpendicular to it ($\theta$).
The quantum effects on these two coordinates are in competition, with the shift towards more shared protons (values of $\delta$ closer to 0) strengthening the hydrogen bond network, and the larger range of angles explored (values of $\theta$ further from 180\degree) weakening it.
Therefore, the ability to describe both of these distributions accurately is closely related to the ability to describe the delicate balance of the competing quantum effects in liquid water. All the geometric properties obtained from our 1-replica contraction simulation were identical to those from simulations with contraction to 6 replicas, confirming that centroid contraction has already converged to the result that would be obtained from a full 32 bead PIMD simulation.

\begin{figure}[h]
\includegraphics[width=\columnwidth]{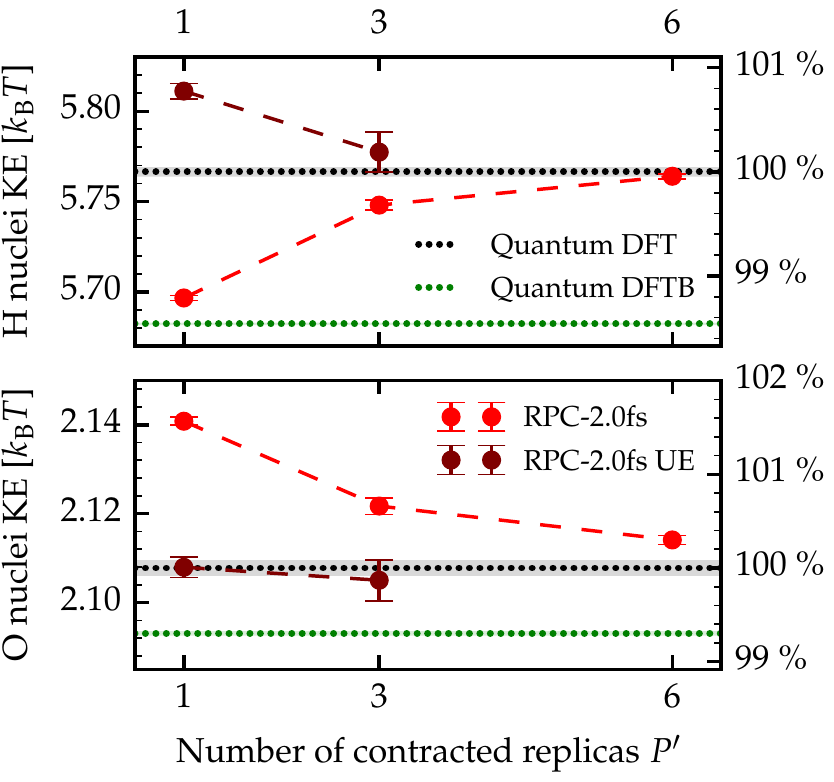}
\caption{
Convergence of the hydrogen (top panel) and oxygen (bottom panel) nuclei quantum kinetic energy in liquid water with respect to the number of contracted replicas $P'$.
Kinetic energies from trajectories run with a fixed total number of replicas $P$=32 and an increasing number of contracted replicas $P'$=1,3,6 and multiple time stepping ($M$=4, $\Delta t$=2.0~fs) are shown in red.
The values obtained using the uncontracted quantum kinetic energy estimator (UE) evaluated on geometries from the $P'$=1 and $P'$=3 trajectories are shown in dark red.
The quantum DFT results were obtained from a 6-replica PIGLET simulation, while the quantum DFTB results were obtained using a full 32-replica PIMD simulation.
The axis on the right shows the percentage of the total kinetic energy obtained relative to the quantum DFT result.
The classical contribution to the kinetic energy for each nucleus is 1.5 $k_{\rm{B}}T$.
Statistical error estimates are indicated using error bars and shading.
\label{fig:water-ke}
}
\end{figure}

Competing quantum effects are also known to manifest in the quantum kinetic energy of liquid water. Hence, finally we consider the convergence of the quantum kinetic energy of H and O nuclei in liquid water, shown in Fig.~\ref{fig:water-ke}. Again, systematic convergence to the quantum DFT result is obtained with an error even with a contraction to 1 replica being within 1.2~\%, or 0.07~$k_{\rm{B}}T$ of the exact result for the H nuclei, which becomes indistinguishable from the statistical error bars of the exact result (below 0.05~\%) with a 6-replica contraction. Using UEs on the centroid contracted trajectory reduces the differences in the quantum kinetic energy from the benchmark to 0.74~\% for H nuclei and 0.04~\% for O nuclei, within the statistical error bar of the exact result of our 100~ps simulations.

\section{Discussion and Conclusions}

In summary, our results show that for both liquid water and the reactive protonated and deprotonated dimer systems, our AI-RPC scheme for DFT, using an SCC-DFTB reference force, is able to obtain quantitative accuracy for a wide range of structural and energetic properties using contraction to the centroid.
Indeed, to distinguish the 1-replica (centroid) contraction results from the statistical error bars of the exact quantum DFT results required trajectories of over 100 ps, which are much longer than those typically performed for AIMD simulations (in total in this study we performed over 3 ns of PIMD and AI-RPC for the gas phase systems and over 300 ps for the liquid water system). When compared to the impact of other subtle factors which can affect the results of AIMD simulations, such as basis set choice~\cite{Ma2012/10.1063/1.4736712} and finite system size effects, such small discrepancies are likely to be unnoticeable in practical applications.

A centroid contracted AI-RPC simulation provides considerable computational speed-ups. Combined with an MTS scheme, a 1-replica contraction simulation reduces the number of DFT electronic structure calculations to evolve 2~fs in time from 128 (32 replicas and 4 time steps) to 1.
This speed-up of two orders of magnitude would only be attainable if the cost of DFTB was truly negligible compared to DFT.
In practice, running the current implementation on a single 16-core node (2x Intel Xeon E5-2670), our centroid contracted simulation of the 64 molecule liquid water system gave 5.6~ps of dynamics per day, which is exactly the same as we obtained from a standard classical AIMD simulation (without MTS).
This represents a 35$\times$ speed-up over a full 32-replica DFT simulation on the same hardware.
Even greater performance gains can be obtained if the difference in the computational cost of the two electronic structure methods was increased.
This would be the case with a faster DFTB implementation or when the DFT calculations become more expensive, for example with a larger basis set.
Additionally, the protonated and deprotonated dimer systems demonstrated that DFTB can act as a good reference potential for the hybrid B3LYP functional, where due to the higher computational cost of hybrid functionals, the speed-ups for condensed phase systems would be nearer the theoretical maximum.
The reason our AI-RPC approach fared well for both a hybrid (B3LYP) and standard GGA (revPBE) exchange correlation functional is that, as explained in Sec.~\ref{subsec:reference}, the reference potential simply needs to capture the broad essence of the full potential to leave a smooth difference potential. Indeed, since the difference of SCC-DFTB from any given density functional is so large compared to the variation among them, it should be an equally good reference for most functionals, making this combination widely applicable.
However, ongoing improvements to DFTB's ability to treat intermolecular interactions ~\cite{Hu2007/10.1021/jp070308d,Gaus2011/10.1021/ct100684s} offer the opportunity to bring it closer to DFT methods and allow even larger time steps and more aggressive contractions to be achieved.

If even tighter accuracy is required, our results demonstrate that AI-RPC exhibits systematic convergence with the number of contracted beads and that uncontracted estimators provide a way to increase the accuracy of the results by post-processing. Also, by examining the difference between the uncontracted and contracted estimator, one can assess whether convergence has been obtained.

To understand the success of AI-RPC, it is important to note that there is a fundamental difference between the performance of the pure reference potential and a centroid contraction.
In AI-RPC the centroid positions are still determined entirely by the full DFT forces and only the higher normal modes of the ring polymers are coupled to the reference forces.
Hence it is instructive to consider the most extreme case, where the reference forces are zero.
This case corresponds to moving under the exact centroid force and then averaging the observables over the free ring polymer distribution (averaged over the free particle position uncertainty).
For interacting systems, this free ring polymer distribution can be noticeably perturbed, e.g. in liquid water it is much more confined along the O-H stretch coordinate.
However, the perturbation to the distribution is dominated by very strong forces in the physical potential, such as chemical bonds.
Therefore, the change in the shape of the ring polymer can be accurately obtained using a reference potential which is able to broadly capture these strong forces.
The local ring polymer distribution produced by the reference potential can then be refined using additional full force evaluations by increasing the contraction level.

Our results also suggest several potential directions for future development.
In particular, the success of the uncontracted estimators, which can be recognized to arise as the zeroth-order cumulant in the expansion of the reweighted trajectory~\cite{Ceriotti2011/10.1098/rspa.2011.0413}, indicates that reweighting the configurations generated from contracted simulations to the full uncontracted Hamiltonian could provide an even larger benefit to properties such as the kinetic energy and also allow position based properties to be improved.
In a similar spirit, due to the well defined Hamiltonian, AI-RPC could also enhance recently proposed path integral replica exchange schemes~\cite{Peng2014/10.1021/ct500447r} by allowing exchanges between different contractions levels, removing the current need to change the number of degrees of freedom upon exchange.
If even higher accuracy is required, one could also envisage using AI-RPC to generate trial moves in a hybrid Monte Carlo scheme~\cite{Iftimie2001/10.1063/1.1357793}.
Finally, analogously to MTS, RPC is not limited to two levels of contraction and hence could be combined with an even higher tier {\it ab inito} method~\cite{Kapil2015/10.1063/1.4941091} at a third level to achieve very high accuracy at moderate cost.
For example, a three-level contraction scheme could combine DFTB on all the replicas, DFT on an intermediate number of replicas and MP2 or another correlated method on the centroid.

In conclusion, our AI-RPC approach provides a method to routinely include nuclear quantum effects exactly in the static equilibrium properties of {\it ab initio} molecular dynamics simulations. 
Additionally, this development offers a unique way to reduce the computational cost of {\it ab initio} CMD and RPMD dynamics, enabling previously unfeasible applications.

\begin{acknowledgements}
The authors thank Will Pfalzgraff and Lu Wang for helpful comments and a thorough reading of this manuscript.
We also thank Lu Wang for providing the revPBE-D3 water quantum DFT trajectory.
This material is based upon work supported by the U.S. Department of Energy, Office of Science, Office of Basic Energy Sciences under Award Number DE-SC0014437.
T.E.M also acknowledges support from a Cottrell Scholarship from the Research Corporation for Science Advancement and an Alfred P. Sloan Research fellowship.
\end{acknowledgements}


\begin{thebibliography}{74}%
\makeatletter
\providecommand \@ifxundefined [1]{%
 \@ifx{#1\undefined}
}%
\providecommand \@ifnum [1]{%
 \ifnum #1\expandafter \@firstoftwo
 \else \expandafter \@secondoftwo
 \fi
}%
\providecommand \@ifx [1]{%
 \ifx #1\expandafter \@firstoftwo
 \else \expandafter \@secondoftwo
 \fi
}%
\providecommand \natexlab [1]{#1}%
\providecommand \enquote  [1]{``#1''}%
\providecommand \bibnamefont  [1]{#1}%
\providecommand \bibfnamefont [1]{#1}%
\providecommand \citenamefont [1]{#1}%
\providecommand \href@noop [0]{\@secondoftwo}%
\providecommand \href [0]{\begingroup \@sanitize@url \@href}%
\providecommand \@href[1]{\@@startlink{#1}\@@href}%
\providecommand \@@href[1]{\endgroup#1\@@endlink}%
\providecommand \@sanitize@url [0]{\catcode `\\12\catcode `\$12\catcode
  `\&12\catcode `\#12\catcode `\^12\catcode `\_12\catcode `\%12\relax}%
\providecommand \@@startlink[1]{}%
\providecommand \@@endlink[0]{}%
\providecommand \url  [0]{\begingroup\@sanitize@url \@url }%
\providecommand \@url [1]{\endgroup\@href {#1}{\urlprefix }}%
\providecommand \urlprefix  [0]{URL }%
\providecommand \Eprint [0]{\href }%
\providecommand \doibase [0]{http://dx.doi.org/}%
\providecommand \selectlanguage [0]{\@gobble}%
\providecommand \bibinfo  [0]{\@secondoftwo}%
\providecommand \bibfield  [0]{\@secondoftwo}%
\providecommand \translation [1]{[#1]}%
\providecommand \BibitemOpen [0]{}%
\providecommand \bibitemStop [0]{}%
\providecommand \bibitemNoStop [0]{.\EOS\space}%
\providecommand \EOS [0]{\spacefactor3000\relax}%
\providecommand \BibitemShut  [1]{\csname bibitem#1\endcsname}%
\let\auto@bib@innerbib\@empty
\bibitem [{\citenamefont {Parrinello}\ and\ \citenamefont
  {Rahman}(1984)}]{Parrinello1984/10.1063/1.446740}%
  \BibitemOpen
  \bibfield  {author} {\bibinfo {author} {\bibfnamefont {M.}~\bibnamefont
  {Parrinello}}\ and\ \bibinfo {author} {\bibfnamefont {A.}~\bibnamefont
  {Rahman}},\ }\href {\doibase 10.1063/1.446740} {\bibfield  {journal}
  {\bibinfo  {journal} {J. Chem. Phys.}\ }\textbf {\bibinfo {volume} {80}},\
  \bibinfo {pages} {860} (\bibinfo {year} {1984})}\BibitemShut {NoStop}%
\bibitem [{\citenamefont {Chandler}\ and\ \citenamefont
  {Wolynes}(1981)}]{Chandler1981/10.1063/1.441588}%
  \BibitemOpen
  \bibfield  {author} {\bibinfo {author} {\bibfnamefont {D.}~\bibnamefont
  {Chandler}}\ and\ \bibinfo {author} {\bibfnamefont {P.~G.}\ \bibnamefont
  {Wolynes}},\ }\href {\doibase 10.1063/1.441588} {\bibfield  {journal}
  {\bibinfo  {journal} {J. Chem. Phys.}\ }\textbf {\bibinfo {volume} {74}},\
  \bibinfo {pages} {4078} (\bibinfo {year} {1981})}\BibitemShut {NoStop}%
\bibitem [{\citenamefont {Feynman}\ and\ \citenamefont
  {Hibbs}(1965)}]{Feynman1965}%
  \BibitemOpen
  \bibfield  {author} {\bibinfo {author} {\bibfnamefont {R.~P.}\ \bibnamefont
  {Feynman}}\ and\ \bibinfo {author} {\bibfnamefont {A.~R.}\ \bibnamefont
  {Hibbs}},\ }\href@noop {} {\emph {\bibinfo {title} {{Quantum mechanics and
  path integrals}}}}\ (\bibinfo  {publisher} {McGraw-Hill New York},\ \bibinfo
  {year} {1965})\BibitemShut {NoStop}%
\bibitem [{\citenamefont {Marx}\ and\ \citenamefont
  {Parrinello}(1994)}]{Marx1994/10.1007/BF01312185}%
  \BibitemOpen
  \bibfield  {author} {\bibinfo {author} {\bibfnamefont {D.}~\bibnamefont
  {Marx}}\ and\ \bibinfo {author} {\bibfnamefont {M.}~\bibnamefont
  {Parrinello}},\ }\href {\doibase 10.1007/BF01312185} {\bibfield  {journal}
  {\bibinfo  {journal} {Zeitschrift fur Phys. B Condens. Matter}\ }\textbf
  {\bibinfo {volume} {95}},\ \bibinfo {pages} {143} (\bibinfo {year}
  {1994})}\BibitemShut {NoStop}%
\bibitem [{\citenamefont {Marx}\ and\ \citenamefont
  {Parrinello}(1996)}]{Marx1996/10.1063/1.471221}%
  \BibitemOpen
  \bibfield  {author} {\bibinfo {author} {\bibfnamefont {D.}~\bibnamefont
  {Marx}}\ and\ \bibinfo {author} {\bibfnamefont {M.}~\bibnamefont
  {Parrinello}},\ }\href {\doibase 10.1063/1.471221} {\bibfield  {journal}
  {\bibinfo  {journal} {J. Chem. Phys.}\ }\textbf {\bibinfo {volume} {104}},\
  \bibinfo {pages} {4077} (\bibinfo {year} {1996})}\BibitemShut {NoStop}%
\bibitem [{\citenamefont {Tuckerman}\ \emph {et~al.}(1996)\citenamefont
  {Tuckerman}, \citenamefont {Marx}, \citenamefont {Klein},\ and\ \citenamefont
  {Parrinello}}]{Tuckerman1996/10.1063/1.471771}%
  \BibitemOpen
  \bibfield  {author} {\bibinfo {author} {\bibfnamefont {M.~E.}\ \bibnamefont
  {Tuckerman}}, \bibinfo {author} {\bibfnamefont {D.}~\bibnamefont {Marx}},
  \bibinfo {author} {\bibfnamefont {M.~L.}\ \bibnamefont {Klein}}, \ and\
  \bibinfo {author} {\bibfnamefont {M.}~\bibnamefont {Parrinello}},\ }\href
  {\doibase 10.1063/1.471771} {\bibfield  {journal} {\bibinfo  {journal} {J.
  Chem. Phys.}\ }\textbf {\bibinfo {volume} {104}},\ \bibinfo {pages} {5579}
  (\bibinfo {year} {1996})}\BibitemShut {NoStop}%
\bibitem [{\citenamefont {Liu}\ \emph {et~al.}(2009)\citenamefont {Liu},
  \citenamefont {Miller}, \citenamefont {Paesani}, \citenamefont {Zhang},\ and\
  \citenamefont {Case}}]{Liu2009/10.1063/1.3254372}%
  \BibitemOpen
  \bibfield  {author} {\bibinfo {author} {\bibfnamefont {J.}~\bibnamefont
  {Liu}}, \bibinfo {author} {\bibfnamefont {W.~H.}\ \bibnamefont {Miller}},
  \bibinfo {author} {\bibfnamefont {F.}~\bibnamefont {Paesani}}, \bibinfo
  {author} {\bibfnamefont {W.}~\bibnamefont {Zhang}}, \ and\ \bibinfo {author}
  {\bibfnamefont {D.~A.}\ \bibnamefont {Case}},\ }\href {\doibase
  10.1063/1.3254372} {\bibfield  {journal} {\bibinfo  {journal} {J. Chem.
  Phys.}\ }\textbf {\bibinfo {volume} {131}},\ \bibinfo {pages} {164509}
  (\bibinfo {year} {2009})}\BibitemShut {NoStop}%
\bibitem [{\citenamefont {Habershon}, \citenamefont {Braams},\ and\
  \citenamefont {Manolopoulos}(2007)}]{Habershon2007/10.1063/1.2786451}%
  \BibitemOpen
  \bibfield  {author} {\bibinfo {author} {\bibfnamefont {S.}~\bibnamefont
  {Habershon}}, \bibinfo {author} {\bibfnamefont {B.~J.}\ \bibnamefont
  {Braams}}, \ and\ \bibinfo {author} {\bibfnamefont {D.~E.}\ \bibnamefont
  {Manolopoulos}},\ }\href {\doibase 10.1063/1.2786451} {\bibfield  {journal}
  {\bibinfo  {journal} {J. Chem. Phys.}\ }\textbf {\bibinfo {volume} {127}},\
  \bibinfo {pages} {174108} (\bibinfo {year} {2007})}\BibitemShut {NoStop}%
\bibitem [{\citenamefont {Rabani}\ \emph {et~al.}(2002)\citenamefont {Rabani},
  \citenamefont {Reichman}, \citenamefont {Krilov},\ and\ \citenamefont
  {Berne}}]{Rabani2002/10.1073/pnas.261540698}%
  \BibitemOpen
  \bibfield  {author} {\bibinfo {author} {\bibfnamefont {E.}~\bibnamefont
  {Rabani}}, \bibinfo {author} {\bibfnamefont {D.~R.}\ \bibnamefont
  {Reichman}}, \bibinfo {author} {\bibfnamefont {G.}~\bibnamefont {Krilov}}, \
  and\ \bibinfo {author} {\bibfnamefont {B.~J.}\ \bibnamefont {Berne}},\ }\href
  {\doibase 10.1073/pnas.261540698} {\bibfield  {journal} {\bibinfo  {journal}
  {Proc. Natl. Acad. Sci. U. S. A.}\ }\textbf {\bibinfo {volume} {99}},\
  \bibinfo {pages} {1129} (\bibinfo {year} {2002})}\BibitemShut {NoStop}%
\bibitem [{\citenamefont {Craig}\ and\ \citenamefont
  {Manolopoulos}(2004)}]{Craig2004/10.1063/1.1777575}%
  \BibitemOpen
  \bibfield  {author} {\bibinfo {author} {\bibfnamefont {I.~R.}\ \bibnamefont
  {Craig}}\ and\ \bibinfo {author} {\bibfnamefont {D.~E.}\ \bibnamefont
  {Manolopoulos}},\ }\href {\doibase 10.1063/1.1777575} {\bibfield  {journal}
  {\bibinfo  {journal} {J. Chem. Phys.}\ }\textbf {\bibinfo {volume} {121}},\
  \bibinfo {pages} {3368} (\bibinfo {year} {2004})}\BibitemShut {NoStop}%
\bibitem [{\citenamefont {Habershon}\ \emph {et~al.}(2013)\citenamefont
  {Habershon}, \citenamefont {Manolopoulos}, \citenamefont {Markland},\ and\
  \citenamefont
  {Miller}}]{Habershon2013/10.1146/annurev-physchem-040412-110122}%
  \BibitemOpen
  \bibfield  {author} {\bibinfo {author} {\bibfnamefont {S.}~\bibnamefont
  {Habershon}}, \bibinfo {author} {\bibfnamefont {D.~E.}\ \bibnamefont
  {Manolopoulos}}, \bibinfo {author} {\bibfnamefont {T.~E.}\ \bibnamefont
  {Markland}}, \ and\ \bibinfo {author} {\bibfnamefont {T.~F.}\ \bibnamefont
  {Miller}},\ }\href {\doibase 10.1146/annurev-physchem-040412-110122}
  {\bibfield  {journal} {\bibinfo  {journal} {Annu. Rev. Phys. Chem.}\ }\textbf
  {\bibinfo {volume} {64}},\ \bibinfo {pages} {387} (\bibinfo {year}
  {2013})}\BibitemShut {NoStop}%
\bibitem [{\citenamefont {Cao}\ and\ \citenamefont
  {Voth}(1994)}]{Cao1994/10.1063/1.468400}%
  \BibitemOpen
  \bibfield  {author} {\bibinfo {author} {\bibfnamefont {J.}~\bibnamefont
  {Cao}}\ and\ \bibinfo {author} {\bibfnamefont {G.~A.}\ \bibnamefont {Voth}},\
  }\href {\doibase 10.1063/1.468400} {\bibfield  {journal} {\bibinfo  {journal}
  {J. Chem. Phys.}\ }\textbf {\bibinfo {volume} {101}},\ \bibinfo {pages}
  {6184} (\bibinfo {year} {1994})}\BibitemShut {NoStop}%
\bibitem [{\citenamefont {Hall}\ and\ \citenamefont
  {Berne}(1984)}]{Hall1984/10.1063/1.448112}%
  \BibitemOpen
  \bibfield  {author} {\bibinfo {author} {\bibfnamefont {R.~W.}\ \bibnamefont
  {Hall}}\ and\ \bibinfo {author} {\bibfnamefont {B.~J.}\ \bibnamefont
  {Berne}},\ }\href {\doibase 10.1063/1.448112} {\bibfield  {journal} {\bibinfo
   {journal} {J. Chem. Phys.}\ }\textbf {\bibinfo {volume} {81}},\ \bibinfo
  {pages} {3641} (\bibinfo {year} {1984})}\BibitemShut {NoStop}%
\bibitem [{\citenamefont {Pollock}\ and\ \citenamefont
  {Ceperley}(1984)}]{Pollock1984/10.1103/PhysRevB.30.2555}%
  \BibitemOpen
  \bibfield  {author} {\bibinfo {author} {\bibfnamefont {E.}~\bibnamefont
  {Pollock}}\ and\ \bibinfo {author} {\bibfnamefont {D.}~\bibnamefont
  {Ceperley}},\ }\href {\doibase 10.1103/PhysRevB.30.2555} {\bibfield
  {journal} {\bibinfo  {journal} {Phys. Rev. B}\ }\textbf {\bibinfo {volume}
  {30}},\ \bibinfo {pages} {2555} (\bibinfo {year} {1984})}\BibitemShut
  {NoStop}%
\bibitem [{\citenamefont {Tuckerman}\ \emph {et~al.}(1993)\citenamefont
  {Tuckerman}, \citenamefont {Berne}, \citenamefont {Martyna},\ and\
  \citenamefont {Klein}}]{Tuckerman1993/10.1063/1.465188}%
  \BibitemOpen
  \bibfield  {author} {\bibinfo {author} {\bibfnamefont {M.~E.}\ \bibnamefont
  {Tuckerman}}, \bibinfo {author} {\bibfnamefont {B.~J.}\ \bibnamefont
  {Berne}}, \bibinfo {author} {\bibfnamefont {G.~J.}\ \bibnamefont {Martyna}},
  \ and\ \bibinfo {author} {\bibfnamefont {M.~L.}\ \bibnamefont {Klein}},\
  }\href {\doibase 10.1063/1.465188} {\bibfield  {journal} {\bibinfo  {journal}
  {J. Chem. Phys.}\ }\textbf {\bibinfo {volume} {99}},\ \bibinfo {pages} {2796}
  (\bibinfo {year} {1993})}\BibitemShut {NoStop}%
\bibitem [{\citenamefont {Ceriotti}\ \emph {et~al.}(2010)\citenamefont
  {Ceriotti}, \citenamefont {Parrinello}, \citenamefont {Markland},\ and\
  \citenamefont {Manolopoulos}}]{Ceriotti2010/10.1063/1.3489925}%
  \BibitemOpen
  \bibfield  {author} {\bibinfo {author} {\bibfnamefont {M.}~\bibnamefont
  {Ceriotti}}, \bibinfo {author} {\bibfnamefont {M.}~\bibnamefont
  {Parrinello}}, \bibinfo {author} {\bibfnamefont {T.~E.}\ \bibnamefont
  {Markland}}, \ and\ \bibinfo {author} {\bibfnamefont {D.~E.}\ \bibnamefont
  {Manolopoulos}},\ }\href {\doibase 10.1063/1.3489925} {\bibfield  {journal}
  {\bibinfo  {journal} {J. Chem. Phys.}\ }\textbf {\bibinfo {volume} {133}},\
  \bibinfo {pages} {124104} (\bibinfo {year} {2010})}\BibitemShut {NoStop}%
\bibitem [{\citenamefont {Ceriotti}, \citenamefont {Manolopoulos},\ and\
  \citenamefont {Parrinello}(2011)}]{Ceriotti2011/10.1063/1.3556661}%
  \BibitemOpen
  \bibfield  {author} {\bibinfo {author} {\bibfnamefont {M.}~\bibnamefont
  {Ceriotti}}, \bibinfo {author} {\bibfnamefont {D.~E.}\ \bibnamefont
  {Manolopoulos}}, \ and\ \bibinfo {author} {\bibfnamefont {M.}~\bibnamefont
  {Parrinello}},\ }\href {\doibase 10.1063/1.3556661} {\bibfield  {journal}
  {\bibinfo  {journal} {J. Chem. Phys.}\ }\textbf {\bibinfo {volume} {134}},\
  \bibinfo {pages} {084104} (\bibinfo {year} {2011})}\BibitemShut {NoStop}%
\bibitem [{\citenamefont {Ceriotti}\ and\ \citenamefont
  {Manolopoulos}(2012)}]{Ceriotti2012/10.1103/PhysRevLett.109.100604}%
  \BibitemOpen
  \bibfield  {author} {\bibinfo {author} {\bibfnamefont {M.}~\bibnamefont
  {Ceriotti}}\ and\ \bibinfo {author} {\bibfnamefont {D.~E.}\ \bibnamefont
  {Manolopoulos}},\ }\href {\doibase 10.1103/PhysRevLett.109.100604} {\bibfield
   {journal} {\bibinfo  {journal} {Phys. Rev. Lett.}\ }\textbf {\bibinfo
  {volume} {109}},\ \bibinfo {pages} {100604} (\bibinfo {year}
  {2012})}\BibitemShut {NoStop}%
\bibitem [{\citenamefont {Shiga}\ and\ \citenamefont
  {Shinoda}(2005)}]{Shiga2005/10.1063/1.2035078}%
  \BibitemOpen
  \bibfield  {author} {\bibinfo {author} {\bibfnamefont {M.}~\bibnamefont
  {Shiga}}\ and\ \bibinfo {author} {\bibfnamefont {W.}~\bibnamefont
  {Shinoda}},\ }\href {\doibase 10.1063/1.2035078} {\bibfield  {journal}
  {\bibinfo  {journal} {J. Chem. Phys.}\ }\textbf {\bibinfo {volume} {123}},\
  \bibinfo {pages} {134502} (\bibinfo {year} {2005})}\BibitemShut {NoStop}%
\bibitem [{\citenamefont {Shinoda}\ and\ \citenamefont
  {Shiga}(2005)}]{Shinoda2005/10.1103/PhysRevE.71.041204}%
  \BibitemOpen
  \bibfield  {author} {\bibinfo {author} {\bibfnamefont {W.}~\bibnamefont
  {Shinoda}}\ and\ \bibinfo {author} {\bibfnamefont {M.}~\bibnamefont
  {Shiga}},\ }\href {\doibase 10.1103/PhysRevE.71.041204} {\bibfield  {journal}
  {\bibinfo  {journal} {Phys. Rev. E}\ }\textbf {\bibinfo {volume} {71}},\
  \bibinfo {pages} {041204} (\bibinfo {year} {2005})}\BibitemShut {NoStop}%
\bibitem [{\citenamefont {P{\'{e}}rez}\ and\ \citenamefont
  {Tuckerman}(2011)}]{Perez2011/10.1063/1.3609120}%
  \BibitemOpen
  \bibfield  {author} {\bibinfo {author} {\bibfnamefont {A.}~\bibnamefont
  {P{\'{e}}rez}}\ and\ \bibinfo {author} {\bibfnamefont {M.~E.}\ \bibnamefont
  {Tuckerman}},\ }\href {\doibase 10.1063/1.3609120} {\bibfield  {journal}
  {\bibinfo  {journal} {J. Chem. Phys.}\ }\textbf {\bibinfo {volume} {135}},\
  \bibinfo {pages} {064104} (\bibinfo {year} {2011})}\BibitemShut {NoStop}%
\bibitem [{\citenamefont {Ceriotti}\ \emph {et~al.}(2011)\citenamefont
  {Ceriotti}, \citenamefont {Brain}, \citenamefont {Riordan},\ and\
  \citenamefont {Manolopoulos}}]{Ceriotti2011/10.1098/rspa.2011.0413}%
  \BibitemOpen
  \bibfield  {author} {\bibinfo {author} {\bibfnamefont {M.}~\bibnamefont
  {Ceriotti}}, \bibinfo {author} {\bibfnamefont {G.~A.~R.}\ \bibnamefont
  {Brain}}, \bibinfo {author} {\bibfnamefont {O.}~\bibnamefont {Riordan}}, \
  and\ \bibinfo {author} {\bibfnamefont {D.~E.}\ \bibnamefont {Manolopoulos}},\
  }\href {\doibase 10.1098/rspa.2011.0413} {\bibfield  {journal} {\bibinfo
  {journal} {Proc. R. Soc. A Math. Phys. Eng. Sci.}\ }\textbf {\bibinfo
  {volume} {468}},\ \bibinfo {pages} {2} (\bibinfo {year} {2011})},\ \Eprint
  {http://arxiv.org/abs/1107.1908} {arXiv:1107.1908} \BibitemShut {NoStop}%
\bibitem [{\citenamefont {Wang}, \citenamefont {Ceriotti},\ and\ \citenamefont
  {Markland}(2014)}]{Wang2014/10.1063/1.4894287}%
  \BibitemOpen
  \bibfield  {author} {\bibinfo {author} {\bibfnamefont {L.}~\bibnamefont
  {Wang}}, \bibinfo {author} {\bibfnamefont {M.}~\bibnamefont {Ceriotti}}, \
  and\ \bibinfo {author} {\bibfnamefont {T.~E.}\ \bibnamefont {Markland}},\
  }\href {\doibase 10.1063/1.4894287} {\bibfield  {journal} {\bibinfo
  {journal} {J. Chem. Phys.}\ }\textbf {\bibinfo {volume} {141}},\ \bibinfo
  {pages} {104502} (\bibinfo {year} {2014})}\BibitemShut {NoStop}%
\bibitem [{\citenamefont {Ceriotti}\ \emph {et~al.}(2013)\citenamefont
  {Ceriotti}, \citenamefont {Cuny}, \citenamefont {Parrinello},\ and\
  \citenamefont {Manolopoulos}}]{Ceriotti2013/10.1073/pnas.1308560110}%
  \BibitemOpen
  \bibfield  {author} {\bibinfo {author} {\bibfnamefont {M.}~\bibnamefont
  {Ceriotti}}, \bibinfo {author} {\bibfnamefont {J.}~\bibnamefont {Cuny}},
  \bibinfo {author} {\bibfnamefont {M.}~\bibnamefont {Parrinello}}, \ and\
  \bibinfo {author} {\bibfnamefont {D.~E.}\ \bibnamefont {Manolopoulos}},\
  }\href {\doibase 10.1073/pnas.1308560110} {\bibfield  {journal} {\bibinfo
  {journal} {Proc. Natl. Acad. Sci. U. S. A.}\ }\textbf {\bibinfo {volume}
  {110}},\ \bibinfo {pages} {15591} (\bibinfo {year} {2013})}\BibitemShut
  {NoStop}%
\bibitem [{\citenamefont {Wang}\ \emph {et~al.}(2014)\citenamefont {Wang},
  \citenamefont {Fried}, \citenamefont {Boxer},\ and\ \citenamefont
  {Markland}}]{Wang2014/10.1073/pnas.1417923111}%
  \BibitemOpen
  \bibfield  {author} {\bibinfo {author} {\bibfnamefont {L.}~\bibnamefont
  {Wang}}, \bibinfo {author} {\bibfnamefont {S.~D.}\ \bibnamefont {Fried}},
  \bibinfo {author} {\bibfnamefont {S.~G.}\ \bibnamefont {Boxer}}, \ and\
  \bibinfo {author} {\bibfnamefont {T.~E.}\ \bibnamefont {Markland}},\ }\href
  {\doibase 10.1073/pnas.1417923111} {\bibfield  {journal} {\bibinfo  {journal}
  {Proc. Natl. Acad. Sci.}\ }\textbf {\bibinfo {volume} {111}},\ \bibinfo
  {pages} {18454} (\bibinfo {year} {2014})}\BibitemShut {NoStop}%
\bibitem [{GLE()}]{GLE-input-generator}%
  \BibitemOpen
  \href {https://epfl-cosmo.github.io/gle4md/index.html?page=matrix} {\enquote
  {\bibinfo {title} {{GLE Input generator,
  https://epfl-cosmo.github.io/gle4md/index.html?page=matrix}},}\ }\BibitemShut
  {NoStop}%
\bibitem [{\citenamefont {Ceriotti}\ and\ \citenamefont
  {Markland}(2013)}]{Ceriotti/10.1063/1.4772676}%
  \BibitemOpen
  \bibfield  {author} {\bibinfo {author} {\bibfnamefont {M.}~\bibnamefont
  {Ceriotti}}\ and\ \bibinfo {author} {\bibfnamefont {T.~E.}\ \bibnamefont
  {Markland}},\ }\href {\doibase 10.1063/1.4772676} {\bibfield  {journal}
  {\bibinfo  {journal} {J. Chem. Phys.}\ }\textbf {\bibinfo {volume} {138}},\
  \bibinfo {pages} {014112} (\bibinfo {year} {2013})}\BibitemShut {NoStop}%
\bibitem [{\citenamefont {Markland}\ and\ \citenamefont
  {Manolopoulos}(2008{\natexlab{a}})}]{Markland2008/10.1063/1.2953308}%
  \BibitemOpen
  \bibfield  {author} {\bibinfo {author} {\bibfnamefont {T.~E.}\ \bibnamefont
  {Markland}}\ and\ \bibinfo {author} {\bibfnamefont {D.~E.}\ \bibnamefont
  {Manolopoulos}},\ }\href {\doibase 10.1063/1.2953308} {\bibfield  {journal}
  {\bibinfo  {journal} {J. Chem. Phys.}\ }\textbf {\bibinfo {volume} {129}},\
  \bibinfo {pages} {024105} (\bibinfo {year} {2008}{\natexlab{a}})}\BibitemShut
  {NoStop}%
\bibitem [{\citenamefont {Markland}\ and\ \citenamefont
  {Manolopoulos}(2008{\natexlab{b}})}]{Markland2008/10.1016/j.cplett.2008.09.019}%
  \BibitemOpen
  \bibfield  {author} {\bibinfo {author} {\bibfnamefont {T.~E.}\ \bibnamefont
  {Markland}}\ and\ \bibinfo {author} {\bibfnamefont {D.~E.}\ \bibnamefont
  {Manolopoulos}},\ }\href {\doibase 10.1016/j.cplett.2008.09.019} {\bibfield
  {journal} {\bibinfo  {journal} {Chem. Phys. Lett.}\ }\textbf {\bibinfo
  {volume} {464}},\ \bibinfo {pages} {256} (\bibinfo {year}
  {2008}{\natexlab{b}})}\BibitemShut {NoStop}%
\bibitem [{\citenamefont {Fanourgakis}, \citenamefont {Markland},\ and\
  \citenamefont {Manolopoulos}(2009)}]{Fanourgakis2009/10.1063/1.3216520}%
  \BibitemOpen
  \bibfield  {author} {\bibinfo {author} {\bibfnamefont {G.~S.}\ \bibnamefont
  {Fanourgakis}}, \bibinfo {author} {\bibfnamefont {T.~E.}\ \bibnamefont
  {Markland}}, \ and\ \bibinfo {author} {\bibfnamefont {D.~E.}\ \bibnamefont
  {Manolopoulos}},\ }\href {\doibase 10.1063/1.3216520} {\bibfield  {journal}
  {\bibinfo  {journal} {J. Chem. Phys.}\ }\textbf {\bibinfo {volume} {131}},\
  \bibinfo {pages} {94102} (\bibinfo {year} {2009})}\BibitemShut {NoStop}%
\bibitem [{\citenamefont {Frauenheim}\ \emph {et~al.}(2000)\citenamefont
  {Frauenheim}, \citenamefont {Seifert}, \citenamefont {Elstner}, \citenamefont
  {Hajnal}, \citenamefont {Jungnikel}, \citenamefont {Porezag}, \citenamefont
  {Suhai},\ and\ \citenamefont {Scholz}}]{Frauenheim2000}%
  \BibitemOpen
  \bibfield  {author} {\bibinfo {author} {\bibfnamefont {T.}~\bibnamefont
  {Frauenheim}}, \bibinfo {author} {\bibfnamefont {G.}~\bibnamefont {Seifert}},
  \bibinfo {author} {\bibfnamefont {M.}~\bibnamefont {Elstner}}, \bibinfo
  {author} {\bibfnamefont {Z.}~\bibnamefont {Hajnal}}, \bibinfo {author}
  {\bibfnamefont {G.}~\bibnamefont {Jungnikel}}, \bibinfo {author}
  {\bibfnamefont {D.}~\bibnamefont {Porezag}}, \bibinfo {author} {\bibfnamefont
  {S.}~\bibnamefont {Suhai}}, \ and\ \bibinfo {author} {\bibfnamefont
  {R.}~\bibnamefont {Scholz}},\ }\href@noop {} {\bibfield  {journal} {\bibinfo
  {journal} {Phys. Stat. Sol.}\ }\textbf {\bibinfo {volume} {217}},\ \bibinfo
  {pages} {41} (\bibinfo {year} {2000})}\BibitemShut {NoStop}%
\bibitem [{\citenamefont {Herman}, \citenamefont {Bruskin},\ and\ \citenamefont
  {Berne}(1982)}]{Herman1982/10.1063/1.442815}%
  \BibitemOpen
  \bibfield  {author} {\bibinfo {author} {\bibfnamefont {M.~F.}\ \bibnamefont
  {Herman}}, \bibinfo {author} {\bibfnamefont {E.~J.}\ \bibnamefont {Bruskin}},
  \ and\ \bibinfo {author} {\bibfnamefont {B.~J.}\ \bibnamefont {Berne}},\
  }\href {\doibase 10.1063/1.442815} {\bibfield  {journal} {\bibinfo  {journal}
  {J. Chem. Phys.}\ }\textbf {\bibinfo {volume} {76}},\ \bibinfo {pages} {5150}
  (\bibinfo {year} {1982})}\BibitemShut {NoStop}%
\bibitem [{\citenamefont {Tuckerman}, \citenamefont {Berne},\ and\
  \citenamefont {Martyna}(1992)}]{Tuckerman1992/10.1063/1.463137}%
  \BibitemOpen
  \bibfield  {author} {\bibinfo {author} {\bibfnamefont {M.}~\bibnamefont
  {Tuckerman}}, \bibinfo {author} {\bibfnamefont {B.~J.}\ \bibnamefont
  {Berne}}, \ and\ \bibinfo {author} {\bibfnamefont {G.~J.}\ \bibnamefont
  {Martyna}},\ }\href {\doibase 10.1063/1.463137} {\bibfield  {journal}
  {\bibinfo  {journal} {J. Chem. Phys.}\ }\textbf {\bibinfo {volume} {97}},\
  \bibinfo {pages} {1990} (\bibinfo {year} {1992})}\BibitemShut {NoStop}%
\bibitem [{\citenamefont {Gaus}, \citenamefont {Cui},\ and\ \citenamefont
  {Elstner}(2011)}]{Gaus2011/10.1021/ct100684s}%
  \BibitemOpen
  \bibfield  {author} {\bibinfo {author} {\bibfnamefont {M.}~\bibnamefont
  {Gaus}}, \bibinfo {author} {\bibfnamefont {Q.}~\bibnamefont {Cui}}, \ and\
  \bibinfo {author} {\bibfnamefont {M.}~\bibnamefont {Elstner}},\ }\href
  {\doibase 10.1021/ct100684s} {\bibfield  {journal} {\bibinfo  {journal} {J.
  Chem. Theory Comput.}\ }\textbf {\bibinfo {volume} {7}},\ \bibinfo {pages}
  {931} (\bibinfo {year} {2011})}\BibitemShut {NoStop}%
\bibitem [{\citenamefont {Anglada}, \citenamefont {Junquera},\ and\
  \citenamefont {Soler}(2003)}]{Anglada2003/10.1103/PhysRevE.68.055701}%
  \BibitemOpen
  \bibfield  {author} {\bibinfo {author} {\bibfnamefont {E.}~\bibnamefont
  {Anglada}}, \bibinfo {author} {\bibfnamefont {J.}~\bibnamefont {Junquera}}, \
  and\ \bibinfo {author} {\bibfnamefont {J.}~\bibnamefont {Soler}},\ }\href
  {\doibase 10.1103/PhysRevE.68.055701} {\bibfield  {journal} {\bibinfo
  {journal} {Phys. Rev. E}\ }\textbf {\bibinfo {volume} {68}},\ \bibinfo
  {pages} {55701} (\bibinfo {year} {2003})}\BibitemShut {NoStop}%
\bibitem [{\citenamefont {Guidon}\ \emph {et~al.}(2008)\citenamefont {Guidon},
  \citenamefont {Schiffmann}, \citenamefont {Hutter},\ and\ \citenamefont
  {VandeVondele}}]{Guidon2008/10.1063/1.2931945}%
  \BibitemOpen
  \bibfield  {author} {\bibinfo {author} {\bibfnamefont {M.}~\bibnamefont
  {Guidon}}, \bibinfo {author} {\bibfnamefont {F.}~\bibnamefont {Schiffmann}},
  \bibinfo {author} {\bibfnamefont {J.}~\bibnamefont {Hutter}}, \ and\ \bibinfo
  {author} {\bibfnamefont {J.}~\bibnamefont {VandeVondele}},\ }\href {\doibase
  10.1063/1.2931945} {\bibfield  {journal} {\bibinfo  {journal} {J. Chem.
  Phys.}\ }\textbf {\bibinfo {volume} {128}},\ \bibinfo {pages} {1} (\bibinfo
  {year} {2008})}\BibitemShut {NoStop}%
\bibitem [{\citenamefont {Luehr}, \citenamefont {Markland},\ and\ \citenamefont
  {Mart{\'{\i}}nez}(2014)}]{Luehr2014/10.1063/1.4866176}%
  \BibitemOpen
  \bibfield  {author} {\bibinfo {author} {\bibfnamefont {N.}~\bibnamefont
  {Luehr}}, \bibinfo {author} {\bibfnamefont {T.~E.}\ \bibnamefont {Markland}},
  \ and\ \bibinfo {author} {\bibfnamefont {T.~J.}\ \bibnamefont
  {Mart{\'{\i}}nez}},\ }\href {\doibase 10.1063/1.4866176} {\bibfield
  {journal} {\bibinfo  {journal} {J. Chem. Phys.}\ }\textbf {\bibinfo {volume}
  {140}},\ \bibinfo {pages} {084116} (\bibinfo {year} {2014})}\BibitemShut
  {NoStop}%
\bibitem [{\citenamefont {Geng}(2015)}]{Geng2015/10.1016/j.jcp.2014.12.007}%
  \BibitemOpen
  \bibfield  {author} {\bibinfo {author} {\bibfnamefont {H.~Y.}\ \bibnamefont
  {Geng}},\ }\href {\doibase 10.1016/j.jcp.2014.12.007} {\bibfield  {journal}
  {\bibinfo  {journal} {J. Comput. Phys.}\ }\textbf {\bibinfo {volume} {283}},\
  \bibinfo {pages} {299} (\bibinfo {year} {2015})}\BibitemShut {NoStop}%
\bibitem [{\citenamefont {Steele}(2013)}]{Steele2013/10.1063/1.4812568}%
  \BibitemOpen
  \bibfield  {author} {\bibinfo {author} {\bibfnamefont {R.~P.}\ \bibnamefont
  {Steele}},\ }\href {\doibase 10.1063/1.4812568} {\bibfield  {journal}
  {\bibinfo  {journal} {J. Chem. Phys.}\ }\textbf {\bibinfo {volume} {139}},\
  \bibinfo {pages} {011102} (\bibinfo {year} {2013})}\BibitemShut {NoStop}%
\bibitem [{\citenamefont {Fatehi}\ and\ \citenamefont
  {Steele}(2015)}]{Fatehi2015/10.1021/ct500904x}%
  \BibitemOpen
  \bibfield  {author} {\bibinfo {author} {\bibfnamefont {S.}~\bibnamefont
  {Fatehi}}\ and\ \bibinfo {author} {\bibfnamefont {R.~P.}\ \bibnamefont
  {Steele}},\ }\href {\doibase 10.1021/ct500904x} {\bibfield  {journal}
  {\bibinfo  {journal} {J. Chem. Theory Comput.}\ }\textbf {\bibinfo {volume}
  {11}},\ \bibinfo {pages} {884} (\bibinfo {year} {2015})}\BibitemShut
  {NoStop}%
\bibitem [{\citenamefont {{Del Ben}}, \citenamefont {Hutter},\ and\
  \citenamefont {VandeVondele}(2015)}]{DelBen2015/10.1063/1.4927325}%
  \BibitemOpen
  \bibfield  {author} {\bibinfo {author} {\bibfnamefont {M.}~\bibnamefont {{Del
  Ben}}}, \bibinfo {author} {\bibfnamefont {J.}~\bibnamefont {Hutter}}, \ and\
  \bibinfo {author} {\bibfnamefont {J.}~\bibnamefont {VandeVondele}},\ }\href
  {\doibase 10.1063/1.4927325} {\bibfield  {journal} {\bibinfo  {journal} {J.
  Chem. Phys.}\ }\textbf {\bibinfo {volume} {143}},\ \bibinfo {pages} {054506}
  (\bibinfo {year} {2015})}\BibitemShut {NoStop}%
\bibitem [{\citenamefont {Habershon}, \citenamefont {Markland},\ and\
  \citenamefont {Manolopoulos}(2009)}]{Habershon2009/10.1063/1.3167790}%
  \BibitemOpen
  \bibfield  {author} {\bibinfo {author} {\bibfnamefont {S.}~\bibnamefont
  {Habershon}}, \bibinfo {author} {\bibfnamefont {T.~E.}\ \bibnamefont
  {Markland}}, \ and\ \bibinfo {author} {\bibfnamefont {D.~E.}\ \bibnamefont
  {Manolopoulos}},\ }\href {\doibase 10.1063/1.3167790} {\bibfield  {journal}
  {\bibinfo  {journal} {J. Chem. Phys.}\ }\textbf {\bibinfo {volume} {131}},\
  \bibinfo {pages} {024501} (\bibinfo {year} {2009})}\BibitemShut {NoStop}%
\bibitem [{\citenamefont {Markland}\ \emph {et~al.}(2011)\citenamefont
  {Markland}, \citenamefont {Morrone}, \citenamefont {Berne}, \citenamefont
  {Miyazaki}, \citenamefont {Rabani},\ and\ \citenamefont
  {Reichman}}]{Markland2011/10.1038/nphys1865}%
  \BibitemOpen
  \bibfield  {author} {\bibinfo {author} {\bibfnamefont {T.~E.}\ \bibnamefont
  {Markland}}, \bibinfo {author} {\bibfnamefont {J.~A.}\ \bibnamefont
  {Morrone}}, \bibinfo {author} {\bibfnamefont {B.~J.}\ \bibnamefont {Berne}},
  \bibinfo {author} {\bibfnamefont {K.}~\bibnamefont {Miyazaki}}, \bibinfo
  {author} {\bibfnamefont {E.}~\bibnamefont {Rabani}}, \ and\ \bibinfo {author}
  {\bibfnamefont {D.~R.}\ \bibnamefont {Reichman}},\ }\href {\doibase
  10.1038/nphys1865} {\bibfield  {journal} {\bibinfo  {journal} {Nat. Phys.}\
  }\textbf {\bibinfo {volume} {7}},\ \bibinfo {pages} {134} (\bibinfo {year}
  {2011})}\BibitemShut {NoStop}%
\bibitem [{\citenamefont {Markland}\ \emph {et~al.}(2012)\citenamefont
  {Markland}, \citenamefont {Morrone}, \citenamefont {Miyazaki}, \citenamefont
  {Berne}, \citenamefont {Reichman},\ and\ \citenamefont
  {Rabani}}]{Markland2012/10.1063/1.3684881}%
  \BibitemOpen
  \bibfield  {author} {\bibinfo {author} {\bibfnamefont {T.~E.}\ \bibnamefont
  {Markland}}, \bibinfo {author} {\bibfnamefont {J.~A.}\ \bibnamefont
  {Morrone}}, \bibinfo {author} {\bibfnamefont {K.}~\bibnamefont {Miyazaki}},
  \bibinfo {author} {\bibfnamefont {B.~J.}\ \bibnamefont {Berne}}, \bibinfo
  {author} {\bibfnamefont {D.~R.}\ \bibnamefont {Reichman}}, \ and\ \bibinfo
  {author} {\bibfnamefont {E.}~\bibnamefont {Rabani}},\ }\href {\doibase
  10.1063/1.3684881} {\bibfield  {journal} {\bibinfo  {journal} {J. Chem.
  Phys.}\ }\textbf {\bibinfo {volume} {136}},\ \bibinfo {pages} {074511}
  (\bibinfo {year} {2012})},\ \Eprint {http://arxiv.org/abs/1111.4191}
  {arXiv:1111.4191} \BibitemShut {NoStop}%
\bibitem [{\citenamefont {Markland}\ and\ \citenamefont
  {Berne}(2012)}]{Markland2012/10.1073/pnas.1203365109}%
  \BibitemOpen
  \bibfield  {author} {\bibinfo {author} {\bibfnamefont {T.~E.}\ \bibnamefont
  {Markland}}\ and\ \bibinfo {author} {\bibfnamefont {B.~J.}\ \bibnamefont
  {Berne}},\ }\href {\doibase 10.1073/pnas.1203365109} {\bibfield  {journal}
  {\bibinfo  {journal} {Proc. Natl. Acad. Sci. U. S. A.}\ }\textbf {\bibinfo
  {volume} {109}},\ \bibinfo {pages} {7988} (\bibinfo {year}
  {2012})}\BibitemShut {NoStop}%
\bibitem [{\citenamefont {Han}\ \emph {et~al.}(2007)\citenamefont {Han},
  \citenamefont {Deng}, \citenamefont {Glimm},\ and\ \citenamefont
  {Martyna}}]{Han2007/10.1016/j.cpc.2006.10.005}%
  \BibitemOpen
  \bibfield  {author} {\bibinfo {author} {\bibfnamefont {G.}~\bibnamefont
  {Han}}, \bibinfo {author} {\bibfnamefont {Y.}~\bibnamefont {Deng}}, \bibinfo
  {author} {\bibfnamefont {J.}~\bibnamefont {Glimm}}, \ and\ \bibinfo {author}
  {\bibfnamefont {G.}~\bibnamefont {Martyna}},\ }\href {\doibase
  10.1016/j.cpc.2006.10.005} {\bibfield  {journal} {\bibinfo  {journal}
  {Comput. Phys. Commun.}\ }\textbf {\bibinfo {volume} {176}},\ \bibinfo
  {pages} {271} (\bibinfo {year} {2007})}\BibitemShut {NoStop}%
\bibitem [{\citenamefont {Hone}, \citenamefont {Rossky},\ and\ \citenamefont
  {Voth}(2006)}]{Hone2006/10.1063/1.2186636}%
  \BibitemOpen
  \bibfield  {author} {\bibinfo {author} {\bibfnamefont {T.~D.}\ \bibnamefont
  {Hone}}, \bibinfo {author} {\bibfnamefont {P.~J.}\ \bibnamefont {Rossky}}, \
  and\ \bibinfo {author} {\bibfnamefont {G.~A.}\ \bibnamefont {Voth}},\ }\href
  {\doibase 10.1063/1.2186636} {\bibfield  {journal} {\bibinfo  {journal} {J.
  Chem. Phys.}\ }\textbf {\bibinfo {volume} {124}},\ \bibinfo {pages} {154103}
  (\bibinfo {year} {2006})}\BibitemShut {NoStop}%
\bibitem [{\citenamefont {Morrone}\ \emph {et~al.}(2011)\citenamefont
  {Morrone}, \citenamefont {Markland}, \citenamefont {Ceriotti},\ and\
  \citenamefont {Berne}}]{Morrone2011/10.1063/1.3518369}%
  \BibitemOpen
  \bibfield  {author} {\bibinfo {author} {\bibfnamefont {J.~A.}\ \bibnamefont
  {Morrone}}, \bibinfo {author} {\bibfnamefont {T.~E.}\ \bibnamefont
  {Markland}}, \bibinfo {author} {\bibfnamefont {M.}~\bibnamefont {Ceriotti}},
  \ and\ \bibinfo {author} {\bibfnamefont {B.~J.}\ \bibnamefont {Berne}},\
  }\href {\doibase 10.1063/1.3518369} {\bibfield  {journal} {\bibinfo
  {journal} {J. Chem. Phys.}\ }\textbf {\bibinfo {volume} {134}},\ \bibinfo
  {pages} {014103} (\bibinfo {year} {2011})}\BibitemShut {NoStop}%
\bibitem [{\citenamefont {Rossi}, \citenamefont {Ceriotti},\ and\ \citenamefont
  {Manolopoulos}(2014)}]{Rossi2014/10.1063/1.4883861}%
  \BibitemOpen
  \bibfield  {author} {\bibinfo {author} {\bibfnamefont {M.}~\bibnamefont
  {Rossi}}, \bibinfo {author} {\bibfnamefont {M.}~\bibnamefont {Ceriotti}}, \
  and\ \bibinfo {author} {\bibfnamefont {D.~E.}\ \bibnamefont {Manolopoulos}},\
  }\href {\doibase 10.1063/1.4883861} {\bibfield  {journal} {\bibinfo
  {journal} {J. Chem. Phys.}\ }\textbf {\bibinfo {volume} {140}},\ \bibinfo
  {pages} {234116} (\bibinfo {year} {2014})},\ \Eprint
  {http://arxiv.org/abs/1406.1074v1} {arXiv:1406.1074v1} \BibitemShut {NoStop}%
\bibitem [{\citenamefont {Kapil}, \citenamefont {VandeVondele},\ and\
  \citenamefont {Ceriotti}(2016)}]{Kapil2015/10.1063/1.4941091}%
  \BibitemOpen
  \bibfield  {author} {\bibinfo {author} {\bibfnamefont {V.}~\bibnamefont
  {Kapil}}, \bibinfo {author} {\bibfnamefont {J.}~\bibnamefont {VandeVondele}},
  \ and\ \bibinfo {author} {\bibfnamefont {M.}~\bibnamefont {Ceriotti}},\
  }\href {\doibase 10.1063/1.4941091} {\bibfield  {journal} {\bibinfo
  {journal} {J. Chem. Phys.}\ }\textbf {\bibinfo {volume} {144}},\ \bibinfo
  {pages} {054111} (\bibinfo {year} {2016})},\ \Eprint
  {http://arxiv.org/abs/1512.00176} {arXiv:1512.00176} \BibitemShut {NoStop}%
\bibitem [{\citenamefont {Bussi}, \citenamefont {Donadio},\ and\ \citenamefont
  {Parrinello}(2007)}]{Bussi2007/10.1063/1.2408420}%
  \BibitemOpen
  \bibfield  {author} {\bibinfo {author} {\bibfnamefont {G.}~\bibnamefont
  {Bussi}}, \bibinfo {author} {\bibfnamefont {D.}~\bibnamefont {Donadio}}, \
  and\ \bibinfo {author} {\bibfnamefont {M.}~\bibnamefont {Parrinello}},\
  }\href {\doibase 10.1063/1.2408420} {\bibfield  {journal} {\bibinfo
  {journal} {J. Chem. Phys.}\ }\textbf {\bibinfo {volume} {126}},\ \bibinfo
  {pages} {014101} (\bibinfo {year} {2007})}\BibitemShut {NoStop}%
\bibitem [{\citenamefont {Ceriotti}, \citenamefont {More},\ and\ \citenamefont
  {Manolopoulos}(2013)}]{Ceriotti2013/10.1016/j.cpc.2013.10.027}%
  \BibitemOpen
  \bibfield  {author} {\bibinfo {author} {\bibfnamefont {M.}~\bibnamefont
  {Ceriotti}}, \bibinfo {author} {\bibfnamefont {J.}~\bibnamefont {More}}, \
  and\ \bibinfo {author} {\bibfnamefont {D.~E.}\ \bibnamefont {Manolopoulos}},\
  }\href {\doibase 10.1016/j.cpc.2013.10.027} {\bibfield  {journal} {\bibinfo
  {journal} {Comput. Phys. Commun.}\ }\textbf {\bibinfo {volume} {185}},\
  \bibinfo {pages} {1019} (\bibinfo {year} {2013})}\BibitemShut {NoStop}%
\bibitem [{\citenamefont {Becke}(1993)}]{Becke1993/10.1063/1.464913}%
  \BibitemOpen
  \bibfield  {author} {\bibinfo {author} {\bibfnamefont {A.~D.}\ \bibnamefont
  {Becke}},\ }\href {\doibase 10.1063/1.464913} {\bibfield  {journal} {\bibinfo
   {journal} {J. Chem. Phys.}\ }\textbf {\bibinfo {volume} {98}},\ \bibinfo
  {pages} {5648} (\bibinfo {year} {1993})}\BibitemShut {NoStop}%
\bibitem [{\citenamefont {Frisch}\ \emph {et~al.}(2009)\citenamefont {Frisch},
  \citenamefont {Trucks}, \citenamefont {Schlegel}, \citenamefont {Scuseria},
  \citenamefont {Robb}, \citenamefont {Cheeseman}, \citenamefont {Scalmani},
  \citenamefont {Barone}, \citenamefont {Mennucci}, \citenamefont {Petersson},
  \citenamefont {Nakatsuji}, \citenamefont {Caricato}, \citenamefont {Li},
  \citenamefont {Hratchian}, \citenamefont {Izmaylov}, \citenamefont {Bloino},
  \citenamefont {Zheng}, \citenamefont {Sonnenberg}, \citenamefont {Hada},
  \citenamefont {Ehara}, \citenamefont {Toyota}, \citenamefont {Fukuda},
  \citenamefont {Hasegawa}, \citenamefont {Ishida}, \citenamefont {Nakajima},
  \citenamefont {Honda}, \citenamefont {Kitao}, \citenamefont {Nakai},
  \citenamefont {Vreven}, \citenamefont {Montgomery}, \citenamefont {Peralta},
  \citenamefont {Ogliaro}, \citenamefont {Bearpark}, \citenamefont {Heyd},
  \citenamefont {Brothers}, \citenamefont {Kudin}, \citenamefont {Staroverov},
  \citenamefont {Kobayashi}, \citenamefont {Normand}, \citenamefont
  {Raghavachari}, \citenamefont {Rendell}, \citenamefont {Burant},
  \citenamefont {Iyengar}, \citenamefont {Tomasi}, \citenamefont {Cossi},
  \citenamefont {Rega}, \citenamefont {Millam}, \citenamefont {Klene},
  \citenamefont {Knox}, \citenamefont {Cross}, \citenamefont {Bakken},
  \citenamefont {Adamo}, \citenamefont {Jaramillo}, \citenamefont {Gomperts},
  \citenamefont {Stratmann}, \citenamefont {Yazyev}, \citenamefont {Austin},
  \citenamefont {Cammi}, \citenamefont {Pomelli}, \citenamefont {Ochterski},
  \citenamefont {Martin}, \citenamefont {Morokuma}, \citenamefont {Zakrzewski},
  \citenamefont {Voth}, \citenamefont {Salvador}, \citenamefont {Dannenberg},
  \citenamefont {Dapprich}, \citenamefont {Daniels}, \citenamefont {Farkas},
  \citenamefont {Foresman}, \citenamefont {Ortiz}, \citenamefont {Cioslowski},\
  and\ \citenamefont {Fox}}]{g09}%
  \BibitemOpen
  \bibfield  {author} {\bibinfo {author} {\bibfnamefont {M.~J.}\ \bibnamefont
  {Frisch}}, \bibinfo {author} {\bibfnamefont {G.~W.}\ \bibnamefont {Trucks}},
  \bibinfo {author} {\bibfnamefont {H.~B.}\ \bibnamefont {Schlegel}}, \bibinfo
  {author} {\bibfnamefont {G.~E.}\ \bibnamefont {Scuseria}}, \bibinfo {author}
  {\bibfnamefont {M.~A.}\ \bibnamefont {Robb}}, \bibinfo {author}
  {\bibfnamefont {J.~R.}\ \bibnamefont {Cheeseman}}, \bibinfo {author}
  {\bibfnamefont {G.}~\bibnamefont {Scalmani}}, \bibinfo {author}
  {\bibfnamefont {V.}~\bibnamefont {Barone}}, \bibinfo {author} {\bibfnamefont
  {B.}~\bibnamefont {Mennucci}}, \bibinfo {author} {\bibfnamefont {G.~A.}\
  \bibnamefont {Petersson}}, \bibinfo {author} {\bibfnamefont {H.}~\bibnamefont
  {Nakatsuji}}, \bibinfo {author} {\bibfnamefont {M.}~\bibnamefont {Caricato}},
  \bibinfo {author} {\bibfnamefont {X.}~\bibnamefont {Li}}, \bibinfo {author}
  {\bibfnamefont {H.~P.}\ \bibnamefont {Hratchian}}, \bibinfo {author}
  {\bibfnamefont {A.~F.}\ \bibnamefont {Izmaylov}}, \bibinfo {author}
  {\bibfnamefont {J.}~\bibnamefont {Bloino}}, \bibinfo {author} {\bibfnamefont
  {G.}~\bibnamefont {Zheng}}, \bibinfo {author} {\bibfnamefont {J.~L.}\
  \bibnamefont {Sonnenberg}}, \bibinfo {author} {\bibfnamefont
  {M.}~\bibnamefont {Hada}}, \bibinfo {author} {\bibfnamefont {M.}~\bibnamefont
  {Ehara}}, \bibinfo {author} {\bibfnamefont {K.}~\bibnamefont {Toyota}},
  \bibinfo {author} {\bibfnamefont {R.}~\bibnamefont {Fukuda}}, \bibinfo
  {author} {\bibfnamefont {J.}~\bibnamefont {Hasegawa}}, \bibinfo {author}
  {\bibfnamefont {M.}~\bibnamefont {Ishida}}, \bibinfo {author} {\bibfnamefont
  {T.}~\bibnamefont {Nakajima}}, \bibinfo {author} {\bibfnamefont
  {Y.}~\bibnamefont {Honda}}, \bibinfo {author} {\bibfnamefont
  {O.}~\bibnamefont {Kitao}}, \bibinfo {author} {\bibfnamefont
  {H.}~\bibnamefont {Nakai}}, \bibinfo {author} {\bibfnamefont
  {T.}~\bibnamefont {Vreven}}, \bibinfo {author} {\bibfnamefont {J.~A.}\
  \bibnamefont {Montgomery}}, \bibinfo {author} {\bibfnamefont {J.~E.}\
  \bibnamefont {Peralta}}, \bibinfo {author} {\bibfnamefont {F.}~\bibnamefont
  {Ogliaro}}, \bibinfo {author} {\bibfnamefont {M.}~\bibnamefont {Bearpark}},
  \bibinfo {author} {\bibfnamefont {J.~J.}\ \bibnamefont {Heyd}}, \bibinfo
  {author} {\bibfnamefont {E.}~\bibnamefont {Brothers}}, \bibinfo {author}
  {\bibfnamefont {K.~N.}\ \bibnamefont {Kudin}}, \bibinfo {author}
  {\bibfnamefont {V.~N.}\ \bibnamefont {Staroverov}}, \bibinfo {author}
  {\bibfnamefont {R.}~\bibnamefont {Kobayashi}}, \bibinfo {author}
  {\bibfnamefont {J.}~\bibnamefont {Normand}}, \bibinfo {author} {\bibfnamefont
  {K.}~\bibnamefont {Raghavachari}}, \bibinfo {author} {\bibfnamefont
  {A.}~\bibnamefont {Rendell}}, \bibinfo {author} {\bibfnamefont {J.~C.}\
  \bibnamefont {Burant}}, \bibinfo {author} {\bibfnamefont {S.~S.}\
  \bibnamefont {Iyengar}}, \bibinfo {author} {\bibfnamefont {J.}~\bibnamefont
  {Tomasi}}, \bibinfo {author} {\bibfnamefont {M.}~\bibnamefont {Cossi}},
  \bibinfo {author} {\bibfnamefont {N.}~\bibnamefont {Rega}}, \bibinfo {author}
  {\bibfnamefont {J.~M.}\ \bibnamefont {Millam}}, \bibinfo {author}
  {\bibfnamefont {M.}~\bibnamefont {Klene}}, \bibinfo {author} {\bibfnamefont
  {J.~E.}\ \bibnamefont {Knox}}, \bibinfo {author} {\bibfnamefont {J.~B.}\
  \bibnamefont {Cross}}, \bibinfo {author} {\bibfnamefont {V.}~\bibnamefont
  {Bakken}}, \bibinfo {author} {\bibfnamefont {C.}~\bibnamefont {Adamo}},
  \bibinfo {author} {\bibfnamefont {J.}~\bibnamefont {Jaramillo}}, \bibinfo
  {author} {\bibfnamefont {R.}~\bibnamefont {Gomperts}}, \bibinfo {author}
  {\bibfnamefont {R.~E.}\ \bibnamefont {Stratmann}}, \bibinfo {author}
  {\bibfnamefont {O.}~\bibnamefont {Yazyev}}, \bibinfo {author} {\bibfnamefont
  {A.~J.}\ \bibnamefont {Austin}}, \bibinfo {author} {\bibfnamefont
  {R.}~\bibnamefont {Cammi}}, \bibinfo {author} {\bibfnamefont
  {C.}~\bibnamefont {Pomelli}}, \bibinfo {author} {\bibfnamefont {J.~W.}\
  \bibnamefont {Ochterski}}, \bibinfo {author} {\bibfnamefont {R.~L.}\
  \bibnamefont {Martin}}, \bibinfo {author} {\bibfnamefont {K.}~\bibnamefont
  {Morokuma}}, \bibinfo {author} {\bibfnamefont {V.~G.}\ \bibnamefont
  {Zakrzewski}}, \bibinfo {author} {\bibfnamefont {G.~A.}\ \bibnamefont
  {Voth}}, \bibinfo {author} {\bibfnamefont {P.}~\bibnamefont {Salvador}},
  \bibinfo {author} {\bibfnamefont {J.~J.}\ \bibnamefont {Dannenberg}},
  \bibinfo {author} {\bibfnamefont {S.}~\bibnamefont {Dapprich}}, \bibinfo
  {author} {\bibfnamefont {A.~D.}\ \bibnamefont {Daniels}}, \bibinfo {author}
  {\bibnamefont {Farkas}}, \bibinfo {author} {\bibfnamefont {J.~B.}\
  \bibnamefont {Foresman}}, \bibinfo {author} {\bibfnamefont {J.~V.}\
  \bibnamefont {Ortiz}}, \bibinfo {author} {\bibfnamefont {J.}~\bibnamefont
  {Cioslowski}}, \ and\ \bibinfo {author} {\bibfnamefont {D.~J.}\ \bibnamefont
  {Fox}},\ }\href@noop {} {\enquote {\bibinfo {title} {{Gaussian 09, Revision
  C.01}},}\ } (\bibinfo {year} {2009})\BibitemShut {NoStop}%
\bibitem [{\citenamefont {Bahn}\ and\ \citenamefont
  {Jacobsen}(2002)}]{Bahn2002/10.1109/5992.998641}%
  \BibitemOpen
  \bibfield  {author} {\bibinfo {author} {\bibfnamefont {S.}~\bibnamefont
  {Bahn}}\ and\ \bibinfo {author} {\bibfnamefont {K.}~\bibnamefont
  {Jacobsen}},\ }\href {\doibase 10.1109/5992.998641} {\bibfield  {journal}
  {\bibinfo  {journal} {Comput. Sci. Eng.}\ }\textbf {\bibinfo {volume} {4}},\
  \bibinfo {pages} {56} (\bibinfo {year} {2002})}\BibitemShut {NoStop}%
\bibitem [{\citenamefont {Hutter}\ \emph {et~al.}(2014)\citenamefont {Hutter},
  \citenamefont {Iannuzzi}, \citenamefont {Schiffmann},\ and\ \citenamefont
  {VandeVondele}}]{Hutter2014/10.1002/wcms.1159}%
  \BibitemOpen
  \bibfield  {author} {\bibinfo {author} {\bibfnamefont {J.}~\bibnamefont
  {Hutter}}, \bibinfo {author} {\bibfnamefont {M.}~\bibnamefont {Iannuzzi}},
  \bibinfo {author} {\bibfnamefont {F.}~\bibnamefont {Schiffmann}}, \ and\
  \bibinfo {author} {\bibfnamefont {J.}~\bibnamefont {VandeVondele}},\ }\href
  {\doibase 10.1002/wcms.1159} {\bibfield  {journal} {\bibinfo  {journal}
  {Wiley Interdiscip. Rev. Comput. Mol. Sci.}\ }\textbf {\bibinfo {volume}
  {4}},\ \bibinfo {pages} {15} (\bibinfo {year} {2014})}\BibitemShut {NoStop}%
\bibitem [{\citenamefont {VandeVondele}\ \emph {et~al.}(2005)\citenamefont
  {VandeVondele}, \citenamefont {Krack}, \citenamefont {Mohamed}, \citenamefont
  {Parrinello}, \citenamefont {Chassaing},\ and\ \citenamefont
  {Hutter}}]{Vandevondele2005/10.1016/j.cpc.2004.12.014}%
  \BibitemOpen
  \bibfield  {author} {\bibinfo {author} {\bibfnamefont {J.}~\bibnamefont
  {VandeVondele}}, \bibinfo {author} {\bibfnamefont {M.}~\bibnamefont {Krack}},
  \bibinfo {author} {\bibfnamefont {F.}~\bibnamefont {Mohamed}}, \bibinfo
  {author} {\bibfnamefont {M.}~\bibnamefont {Parrinello}}, \bibinfo {author}
  {\bibfnamefont {T.}~\bibnamefont {Chassaing}}, \ and\ \bibinfo {author}
  {\bibfnamefont {J.}~\bibnamefont {Hutter}},\ }\href {\doibase
  10.1016/j.cpc.2004.12.014} {\bibfield  {journal} {\bibinfo  {journal}
  {Comput. Phys. Commun.}\ }\textbf {\bibinfo {volume} {167}},\ \bibinfo
  {pages} {103} (\bibinfo {year} {2005})}\BibitemShut {NoStop}%
\bibitem [{\citenamefont {Perdew}, \citenamefont {Burke},\ and\ \citenamefont
  {Ernzerhof}(1996)}]{Perdew1996/10.1103/PhysRevLett.77.3865}%
  \BibitemOpen
  \bibfield  {author} {\bibinfo {author} {\bibfnamefont {J.}~\bibnamefont
  {Perdew}}, \bibinfo {author} {\bibfnamefont {K.}~\bibnamefont {Burke}}, \
  and\ \bibinfo {author} {\bibfnamefont {M.}~\bibnamefont {Ernzerhof}},\ }\href
  {\doibase 10.1103/PhysRevLett.77.3865} {\bibfield  {journal} {\bibinfo
  {journal} {Phys. Rev. Lett.}\ }\textbf {\bibinfo {volume} {77}},\ \bibinfo
  {pages} {3865} (\bibinfo {year} {1996})}\BibitemShut {NoStop}%
\bibitem [{\citenamefont {Zhang}\ and\ \citenamefont
  {Yang}(1998)}]{Zhang1998/10.1103/PhysRevLett.80.890}%
  \BibitemOpen
  \bibfield  {author} {\bibinfo {author} {\bibfnamefont {Y.}~\bibnamefont
  {Zhang}}\ and\ \bibinfo {author} {\bibfnamefont {W.}~\bibnamefont {Yang}},\
  }\href {\doibase 10.1103/PhysRevLett.80.890} {\bibfield  {journal} {\bibinfo
  {journal} {Phys. Rev. Lett.}\ }\textbf {\bibinfo {volume} {80}},\ \bibinfo
  {pages} {890} (\bibinfo {year} {1998})}\BibitemShut {NoStop}%
\bibitem [{\citenamefont {Grimme}\ \emph {et~al.}(2010)\citenamefont {Grimme},
  \citenamefont {Antony}, \citenamefont {Ehrlich},\ and\ \citenamefont
  {Krieg}}]{Grimme2010/10.1063/1.3382344}%
  \BibitemOpen
  \bibfield  {author} {\bibinfo {author} {\bibfnamefont {S.}~\bibnamefont
  {Grimme}}, \bibinfo {author} {\bibfnamefont {J.}~\bibnamefont {Antony}},
  \bibinfo {author} {\bibfnamefont {S.}~\bibnamefont {Ehrlich}}, \ and\
  \bibinfo {author} {\bibfnamefont {H.}~\bibnamefont {Krieg}},\ }\href
  {\doibase 10.1063/1.3382344} {\bibfield  {journal} {\bibinfo  {journal} {J.
  Chem. Phys.}\ }\textbf {\bibinfo {volume} {132}},\ \bibinfo {pages} {154104}
  (\bibinfo {year} {2010})}\BibitemShut {NoStop}%
\bibitem [{\citenamefont {Goedecker}, \citenamefont {Teter},\ and\
  \citenamefont {Hutter}(1996)}]{Goedecker1996/10.1103/PhysRevB.54.1703}%
  \BibitemOpen
  \bibfield  {author} {\bibinfo {author} {\bibfnamefont {S.}~\bibnamefont
  {Goedecker}}, \bibinfo {author} {\bibfnamefont {M.}~\bibnamefont {Teter}}, \
  and\ \bibinfo {author} {\bibfnamefont {J.}~\bibnamefont {Hutter}},\ }\href
  {\doibase 10.1103/PhysRevB.54.1703} {\bibfield  {journal} {\bibinfo
  {journal} {Phys. Rev. B}\ }\textbf {\bibinfo {volume} {54}},\ \bibinfo
  {pages} {1703} (\bibinfo {year} {1996})}\BibitemShut {NoStop}%
\bibitem [{\citenamefont {Lippert}, \citenamefont {Hutter},\ and\ \citenamefont
  {Parrinello}(1997)}]{Lippert1997/10.1080/002689797170220}%
  \BibitemOpen
  \bibfield  {author} {\bibinfo {author} {\bibfnamefont {G.}~\bibnamefont
  {Lippert}}, \bibinfo {author} {\bibfnamefont {J.}~\bibnamefont {Hutter}}, \
  and\ \bibinfo {author} {\bibfnamefont {M.}~\bibnamefont {Parrinello}},\
  }\href {\doibase 10.1080/002689797170220} {\bibfield  {journal} {\bibinfo
  {journal} {Mol. Phys.}\ }\textbf {\bibinfo {volume} {92}},\ \bibinfo {pages}
  {477} (\bibinfo {year} {1997})}\BibitemShut {NoStop}%
\bibitem [{\citenamefont {VandeVondele}\ and\ \citenamefont
  {Hutter}(2003)}]{VandeVondele2003/10.1063/1.1543154}%
  \BibitemOpen
  \bibfield  {author} {\bibinfo {author} {\bibfnamefont {J.}~\bibnamefont
  {VandeVondele}}\ and\ \bibinfo {author} {\bibfnamefont {J.}~\bibnamefont
  {Hutter}},\ }\href {\doibase 10.1063/1.1543154} {\bibfield  {journal}
  {\bibinfo  {journal} {J. Chem. Phys.}\ }\textbf {\bibinfo {volume} {118}},\
  \bibinfo {pages} {4365} (\bibinfo {year} {2003})}\BibitemShut {NoStop}%
\bibitem [{\citenamefont {Kolafa}(2004)}]{Kolafa2004/10.1002/jcc.10385}%
  \BibitemOpen
  \bibfield  {author} {\bibinfo {author} {\bibfnamefont {J.}~\bibnamefont
  {Kolafa}},\ }\href {\doibase 10.1002/jcc.10385} {\bibfield  {journal}
  {\bibinfo  {journal} {J. Comput. Chem.}\ }\textbf {\bibinfo {volume} {25}},\
  \bibinfo {pages} {335} (\bibinfo {year} {2004})}\BibitemShut {NoStop}%
\bibitem [{\citenamefont {Richters}\ and\ \citenamefont
  {K{\"{u}}hne}(2014)}]{Richters2014/10.1063/1.4869865}%
  \BibitemOpen
  \bibfield  {author} {\bibinfo {author} {\bibfnamefont {D.}~\bibnamefont
  {Richters}}\ and\ \bibinfo {author} {\bibfnamefont {T.~D.}\ \bibnamefont
  {K{\"{u}}hne}},\ }\href {\doibase 10.1063/1.4869865} {\bibfield  {journal}
  {\bibinfo  {journal} {J. Chem. Phys.}\ }\textbf {\bibinfo {volume} {140}},\
  \bibinfo {pages} {134109} (\bibinfo {year} {2014})}\BibitemShut {NoStop}%
\bibitem [{\citenamefont {Young}(2015)}]{Young2012/10.1007/978-3-319-19051-8}%
  \BibitemOpen
  \bibfield  {author} {\bibinfo {author} {\bibfnamefont {P.}~\bibnamefont
  {Young}},\ }\href {\doibase 10.1007/978-3-319-19051-8} {\emph {\bibinfo
  {title} {{Everything You Wanted to Know About Data Analysis and Fitting but
  Were Afraid to Ask}}}},\ SpringerBriefs in Physics\ (\bibinfo  {publisher}
  {Springer International Publishing},\ \bibinfo {year} {2015})\ \Eprint
  {http://arxiv.org/abs/1210.3781} {arXiv:1210.3781} \BibitemShut {NoStop}%
\bibitem [{\citenamefont {Tuckerman}\ \emph {et~al.}(1997)\citenamefont
  {Tuckerman}, \citenamefont {Marx}, \citenamefont {Klein},\ and\ \citenamefont
  {Parrinello}}]{Tuckerman1997/10.1126/science.275.5301.817}%
  \BibitemOpen
  \bibfield  {author} {\bibinfo {author} {\bibfnamefont {M.~E.}\ \bibnamefont
  {Tuckerman}}, \bibinfo {author} {\bibfnamefont {D.}~\bibnamefont {Marx}},
  \bibinfo {author} {\bibfnamefont {M.~L.}\ \bibnamefont {Klein}}, \ and\
  \bibinfo {author} {\bibfnamefont {M.}~\bibnamefont {Parrinello}},\ }\href
  {\doibase 10.1126/science.275.5301.817} {\bibfield  {journal} {\bibinfo
  {journal} {Science}\ }\textbf {\bibinfo {volume} {275}},\ \bibinfo {pages}
  {817} (\bibinfo {year} {1997})}\BibitemShut {NoStop}%
\bibitem [{\citenamefont {Koskinen}\ and\ \citenamefont
  {M{\"{a}}kinen}(2009)}]{Koskinen2009/10.1016/j.commatsci.2009.07.013}%
  \BibitemOpen
  \bibfield  {author} {\bibinfo {author} {\bibfnamefont {P.}~\bibnamefont
  {Koskinen}}\ and\ \bibinfo {author} {\bibfnamefont {V.}~\bibnamefont
  {M{\"{a}}kinen}},\ }\href {\doibase 10.1016/j.commatsci.2009.07.013}
  {\bibfield  {journal} {\bibinfo  {journal} {Comput. Mater. Sci.}\ }\textbf
  {\bibinfo {volume} {47}},\ \bibinfo {pages} {237} (\bibinfo {year} {2009})},\
  \Eprint {http://arxiv.org/abs/0910.5861} {arXiv:0910.5861} \BibitemShut
  {NoStop}%
\bibitem [{\citenamefont {Maupin}, \citenamefont {Aradi},\ and\ \citenamefont
  {Voth}(2010)}]{Maupin2010/10.1021/jp1010555}%
  \BibitemOpen
  \bibfield  {author} {\bibinfo {author} {\bibfnamefont {C.~M.}\ \bibnamefont
  {Maupin}}, \bibinfo {author} {\bibfnamefont {B.}~\bibnamefont {Aradi}}, \
  and\ \bibinfo {author} {\bibfnamefont {G.~A.}\ \bibnamefont {Voth}},\ }\href
  {\doibase 10.1021/jp1010555} {\bibfield  {journal} {\bibinfo  {journal} {J.
  Phys. Chem. B}\ }\textbf {\bibinfo {volume} {114}},\ \bibinfo {pages} {6922}
  (\bibinfo {year} {2010})}\BibitemShut {NoStop}%
\bibitem [{\citenamefont {Goyal}\ \emph {et~al.}(2014)\citenamefont {Goyal},
  \citenamefont {Qian}, \citenamefont {Irle}, \citenamefont {Lu}, \citenamefont
  {Roston}, \citenamefont {Mori}, \citenamefont {Elstner},\ and\ \citenamefont
  {Cui}}]{Goyal2013/10.1021/jp503372v}%
  \BibitemOpen
  \bibfield  {author} {\bibinfo {author} {\bibfnamefont {P.}~\bibnamefont
  {Goyal}}, \bibinfo {author} {\bibfnamefont {H.-J.}\ \bibnamefont {Qian}},
  \bibinfo {author} {\bibfnamefont {S.}~\bibnamefont {Irle}}, \bibinfo {author}
  {\bibfnamefont {X.}~\bibnamefont {Lu}}, \bibinfo {author} {\bibfnamefont
  {D.}~\bibnamefont {Roston}}, \bibinfo {author} {\bibfnamefont
  {T.}~\bibnamefont {Mori}}, \bibinfo {author} {\bibfnamefont {M.}~\bibnamefont
  {Elstner}}, \ and\ \bibinfo {author} {\bibfnamefont {Q.}~\bibnamefont
  {Cui}},\ }\href {\doibase 10.1021/jp503372v} {\bibfield  {journal} {\bibinfo
  {journal} {J. Phys. Chem. B}\ }\textbf {\bibinfo {volume} {118}},\ \bibinfo
  {pages} {11007} (\bibinfo {year} {2014})}\BibitemShut {NoStop}%
\bibitem [{\citenamefont {Hu}\ \emph {et~al.}(2007)\citenamefont {Hu},
  \citenamefont {Lu}, \citenamefont {Elstner}, \citenamefont {Hermans},\ and\
  \citenamefont {Yang}}]{Hu2007/10.1021/jp070308d}%
  \BibitemOpen
  \bibfield  {author} {\bibinfo {author} {\bibfnamefont {H.}~\bibnamefont
  {Hu}}, \bibinfo {author} {\bibfnamefont {Z.}~\bibnamefont {Lu}}, \bibinfo
  {author} {\bibfnamefont {M.}~\bibnamefont {Elstner}}, \bibinfo {author}
  {\bibfnamefont {J.}~\bibnamefont {Hermans}}, \ and\ \bibinfo {author}
  {\bibfnamefont {W.}~\bibnamefont {Yang}},\ }\href {\doibase
  10.1021/jp070308d} {\bibfield  {journal} {\bibinfo  {journal} {J. Phys. Chem.
  A}\ }\textbf {\bibinfo {volume} {111}},\ \bibinfo {pages} {5685} (\bibinfo
  {year} {2007})}\BibitemShut {NoStop}%
\bibitem [{\citenamefont {Ma}, \citenamefont {Zhang},\ and\ \citenamefont
  {Tuckerman}(2012)}]{Ma2012/10.1063/1.4736712}%
  \BibitemOpen
  \bibfield  {author} {\bibinfo {author} {\bibfnamefont {Z.}~\bibnamefont
  {Ma}}, \bibinfo {author} {\bibfnamefont {Y.}~\bibnamefont {Zhang}}, \ and\
  \bibinfo {author} {\bibfnamefont {M.~E.}\ \bibnamefont {Tuckerman}},\ }\href
  {\doibase 10.1063/1.4736712} {\bibfield  {journal} {\bibinfo  {journal} {J.
  Chem. Phys.}\ }\textbf {\bibinfo {volume} {137}},\ \bibinfo {pages} {044506}
  (\bibinfo {year} {2012})}\BibitemShut {NoStop}%
\bibitem [{\citenamefont {Peng}\ \emph {et~al.}(2014)\citenamefont {Peng},
  \citenamefont {Cao}, \citenamefont {Zhou},\ and\ \citenamefont
  {Voth}}]{Peng2014/10.1021/ct500447r}%
  \BibitemOpen
  \bibfield  {author} {\bibinfo {author} {\bibfnamefont {Y.}~\bibnamefont
  {Peng}}, \bibinfo {author} {\bibfnamefont {Z.}~\bibnamefont {Cao}}, \bibinfo
  {author} {\bibfnamefont {R.}~\bibnamefont {Zhou}}, \ and\ \bibinfo {author}
  {\bibfnamefont {G.~A.}\ \bibnamefont {Voth}},\ }\href {\doibase
  10.1021/ct500447r} {\bibfield  {journal} {\bibinfo  {journal} {J. Chem.
  Theory Comput.}\ }\textbf {\bibinfo {volume} {10}},\ \bibinfo {pages} {3634}
  (\bibinfo {year} {2014})}\BibitemShut {NoStop}%
\bibitem [{\citenamefont {Iftimie}\ and\ \citenamefont
  {Schofield}(2001)}]{Iftimie2001/10.1063/1.1357793}%
  \BibitemOpen
  \bibfield  {author} {\bibinfo {author} {\bibfnamefont {R.}~\bibnamefont
  {Iftimie}}\ and\ \bibinfo {author} {\bibfnamefont {J.}~\bibnamefont
  {Schofield}},\ }\href {\doibase 10.1063/1.1357793} {\bibfield  {journal}
  {\bibinfo  {journal} {J. Chem. Phys.}\ }\textbf {\bibinfo {volume} {114}},\
  \bibinfo {pages} {6763} (\bibinfo {year} {2001})}\BibitemShut {NoStop}%
\end{thebibliography}
\end{document}